\documentclass[ 
superscriptaddress,
 amsmath,amssymb,
 aps, 
 physrev,
]{revtex4-2}
\usepackage{graphicx}
\usepackage{dcolumn}
\usepackage{bm}
\usepackage[dvipsnames]{xcolor}
\usepackage{soul}
\usepackage{comment}
\usepackage[utf8]{inputenc}
\usepackage[T1]{fontenc}
\usepackage{mathptmx}
\usepackage{etoolbox}
\usepackage{ulem}
\usepackage{calrsfs}
\DeclareMathAlphabet{\pazocal}{OMS}{zplm}{m}{n}
 \newcommand{\BV}{Brunt-V\"ais\"al\"a\ } 
 \newcommand{\be}{\begin{equation}}  
 \newcommand{\ee}{\end{equation}} 
 \newcommand{\vomega}{\mbox{\boldmath $\omega$}} 
  
\sethlcolor{yellow!20}
\makeatletter
\def\@email#1#2{%
 \endgroup
 \patchcmd{\titleblock@produce}
 {\frontmatter@RRAPformat}
 {\frontmatter@RRAPformat{\produce@RRAP{*#1\href{mailto:#2}{#2}}}\frontmatter@RRAPformat}
 {}{}
}%
\makeatother
\begin{document}


\title{Scaling laws and local enhancements of buoyancy flux in stratified turbulent flows}
%
%
\author{Gyeongseob Song}
\affiliation{CNRS, \'Ecole Centrale de Lyon, INSA Lyon, Universit\'e Claude Bernard Lyon 1, LMFA, UMR5509, F-69134 \'Ecully, France}
\author{Fabio Feraco}
\affiliation{CNRS, \'Ecole Centrale de Lyon, INSA Lyon, Universit\'e Claude Bernard Lyon 1, LMFA, UMR5509, F-69134 \'Ecully, France}
\affiliation{Dipartimento di Fisica, Universit\`a della Calabria, Italy}
\author{Raffaele Marino}
\affiliation{CNRS, \'Ecole Centrale de Lyon, INSA Lyon, Universit\'e Claude Bernard Lyon 1, LMFA, UMR5509, F-69134 \'Ecully, France}
\author{Jorge L. Chau}
\affiliation{Leibniz Institute of Atmospheric Physics, University of Rostock, K\"uhlungsborn, Germany}
\author{Alain Pumir}
\affiliation{French American Center for Theoretical Sciences, CNRS, KITP, University of California, Santa Barbara, CA 93106-4030, USA}
\affiliation{Laboratoire de Physique, Ecole Normale Sup\'erieure de Lyon and CNRS, Lyon France}
\affiliation{Max Planck Institute for Dynamics and Self-Organization, G\"ottinten, D-37077, Germany}
\author{Leonardo Primavera}
\affiliation{Dipartimento di Fisica, Universit\`a della Calabria, Italy}
\author{Annick Pouquet}
\affiliation{National Center for Atmospheric Research, Boulder, CO 80303 USA}
\author{Pablo D. Mininni}
\affiliation{Universidad de Buenos Aires, Facultad de Ciencias Exactas y Naturales, Departamento de Física, Ciudad Universitaria, 1428 Buenos Aires, Argentina,}
\affiliation{CONICET - Universidad de Buenos Aires, Instituto de F\'{\i}sica Interdisciplinaria y Aplicada (INFINA), Ciudad Universitaria, 1428 Buenos Aires, Argentina.}
\author{Duane Rosenberg}
\affiliation{Cooperative Institute for Research in the Atmosphere, Boulder, CO 80305, USA}



\begin{abstract}
In the presence of stratification, turbulent flows exhibit intermittency not only at small scales but also at large scales, 
comparable to that of the mean flow, as observed in the atmosphere and oceans. We study the dynamics of such flows by performing a large parametric exploration using direct numerical simulations of the Boussinesq equations with different types of forcing. We examine two values of the ratio between viscosity and diffusivity---defining the Prandtl number, taken to be 1 and 6---and vary the Froude number ($Fr$) over a range of values of geophysical interest, $0.01 \le Fr \le 1$, here corresponding to a variation in terms of the buoyancy Reynolds number ($R_{IB}$) of $0.06 \le R_{IB} \le 2300$. 
Specifically, we analyze the dependence on $R_{IB}$ of the buoyancy flux, the mixing efficiency, the shear parameters, and the vertical momentum flux. Strongly non-Gaussian tails in the spatio-temporal distribution of the buoyancy flux are observed, with kurtosis as high as $\approx 10^2$, indicating the potential for stratified geophysical flows to be characterized by highly variable transport properties along the direction of gravity, even in the presence of stable stratification. 
This is associated with the long-time intermittent behavior of vertical velocity and temperature at large scale, which produces local turbulence and enhances dissipation and transport in geophysical flows. We present evidence that the skewness of $B_f$ increases with $R_{IB}$,
following a power-law scaling as stratification weakens, and saturates in
the passive-scalar limit, where temperature decouples from the velocity
field. We also show that the domain-averaged $B_f$ exhibits two distinct
trends, scaling logarithmically with $R_{IB}$ and approaching a small
offset as stratification strengthens. A simple model for the temporal evolution of energy and buoyancy flux clearly indicates that it is the energy defect between vertical and potential energy that drives the strong events in buoyancy flux.
This trend directly leads to convective instabilities, the formation of two-dimensional and three-dimensional eddies, and rapid dissipation on the scale of a turnover time, allowing for the energetic cycle to restart---also occurring in bursts. 
\end{abstract}
\maketitle

\section{Introduction}
The atmosphere and oceans are in quasi-geostrophic equilibrium at large scales, resulting from a balance between pressure gradient, gravity, and the Coriolis forces, with zonal flows being the dominant features, together with inertia-gravity waves and localized strong turbulence. Indeed, the large-scale flow can be strongly perturbed locally by wave breaking in the upper layer of the atmosphere \cite{sun_15}, for example in the tropopause \cite{fritts_17, rodriguez_23} and the mesosphere \cite{Becker, chau_20}, or in the ocean in highly localized dissipation events \cite{pearson_18}, {\it e.g.}, in the vicinity of the Hawaiian Ridge \cite{klymak_08} or in the Puerto-Rico Trench \cite{vanharen_16j}. In all cases, these events lead to efficient localized dissipation and mixing, several orders of magnitude higher than the average and which may also differ from that of a passive tracer \cite{sreenivasan_19}. 
Wave breaking in the tropopause is also seen as a source of blocking events, as reviewed recently \cite{lucarini_20}: these events can last for up to several weeks and they may in fact become predictable in a near future \cite{athanasiadis_20}. 

The dynamics of turbulent flows is known to involve interactions between widely separated scales.
In particular, such interactions are at play for coherent structures, such as vortex filaments or current sheets in magnetohydrodynamic, structures which are locally not isotropic, with an aspect ratio close to the ratio of the integral to the dissipative scale.
Indeed, fully developed, homogeneous and isotropic turbulence (HIT) in fluids is known to be intermittent at small scales, with non-Gaussian wings in the statistics of field gradients and of field increments, and with strong localized dissipative structures. 
These were observed early-on with the help of direct numerical simulations \cite{siggia_81, kerr_85}, as well as in the laboratory \cite{douady_91}. They have also been clearly identified and analyzed more recently in high-resolution simulations with Taylor Reynolds numbers up to $\approx 1300$ \cite{ishihara_09, yeung_15, iyer_20, buaria_22}. 
Moreover, structures in the form of fronts are particularly striking in the case of a tracer passively advected by the flow, and the characteristic skewness (the normalized third-order moment) of the scalar gradients is found to be $\mathcal{O}(1)$ \cite{holzer_94,pumir_94,shraiman_00}.

However, evidence has been mounting that, in some cases, the velocity field itself exhibits non-Gaussian statistics, for example in natural space plasmas \cite{marino_12}, as well as in fluids such as in shear layers \cite{pumir_96, barkley_16, pumir_16}, and in stratified turbulence in the presence of forcing (Rorai et al.\ \cite{rorai_14}, and Feraco et al.\ \cite{feraco_18}, hereafter R14 and F18, respectively). Similar behavior has already been observed in the nocturnal planetary boundary layer \cite{mahrt_89, lenschow_94, lenschow_12}, resulting in long intermittent behavior of stratified flows in a certain regime of the governing parameters, as first shown in simulations (see F18) and modeled by \citet{marino_22} (hereafter M22). This phenomenon has been called ``large-scale'' intermittency as it involves large patches of the flow developing these features due to the emergence of powerful vertical velocity drafts acting as a local energy injection mechanism \cite{foldes_25}.
In the present paper, we extend the analysis of large-scale intermittency in purely stratified turbulence in the presence of forcing, considering different Prandtl numbers, and focusing in particular on the vertical buoyancy flux $B_f$ (see the next section for definitions). 
After writing the equations and describing the numerical framework used and the large parametric exploration we performed in \S \ref{S:DEF}, we analyze in \S \ref{S:INT} the trends followed by the skewness and kurtosis of the primitive variables and of the buoyancy flux ($B_f$) as a function of the controlling parameter, {\it i.e.}, the buoyancy Reynolds number $R_{IB}$. In \S\ref{S:LIMIT}, the two asymptotic regimes of the Boussinesq model, ($N\to\infty$) and ($N\to 0$), are examined.
We also show how stratification modulates mixing and the extreme events developing in both the vertical velocity $w$ and in the temperature fluctuations $\theta$ \cite{feraco_21}, and how it can greatly enhance the buoyancy flux using two-dimensional contour plots of its local mean value in \S\ref{S:DISS}. Finally, \S\ref{S:CONCLU} is our conclusion.

\section{Numerical approach} \label{S:DEF}
We derive our results from a set of direct numerical simulations (DNS) of the Navier-Stokes equations using the Boussinesq approximation in which the velocity field ${\bf u}=(u, v, w)$ remains incompressible, $\nabla \cdot {\bf u} = 0$, but density variations are present in conjunction with the buoyancy term. These equations are written as:
\begin{eqnarray} 
\partial_t {\bf u} + ({\bf u} \cdot \nabla) {\bf u} &=& - \nabla p - N\theta {\bf \hat z} + 
\nu \nabla^2 {\bf u} + {\bf F} , \label{momentum} \\
 \partial_t \theta + ({\bf u} \cdot \nabla) \theta &=& Nw + \kappa \nabla^2 \theta \ . \label{scalar}
\end{eqnarray} 
Temperature fluctuations $\theta$ are evaluated relative to a mean temperature profile $\theta_0$; $N = [-(g/\theta_0)(\partial_z {\bar \theta})]^{1/2}$ is the \BV frequency, where $\partial_z {\bar \theta}$ is the imposed background temperature stratification.
Initial conditions consist of zero temperature fluctuations together with random velocity modes centered on a specific Fourier shell at large scale,
or else with a large-scale Taylor-Green vortex. Additionally, the Navier-Stokes equations decoupled from the equation for the temperature (thus, with $N=0$) are integrated to perform simulations of a passive scalar (hereafter $\rho$), reported in the last column of Table \ref{Tab1}.
Two types of forcing, ${\bf F}$, have been used for the velocity field only: a random forcing imposed in the wavenumber shell, $2 \le k^{(PP)}_F \le 3$, and with negligible global helicity, using a Pouquet-Patterson scheme \cite{patterson_78}; and a Taylor-Green forcing injecting energy within the shell $k^{(TG)}_F= \sqrt{3}$ \cite{marino_14} (see also \cite{Sujovolsky} for more details). 
In the above equations, $\nu$ and $\kappa$ are the kinematic viscosity and the thermal diffusivity, respectively, and we take the Prandtl number, defined as $Pr=\nu/\kappa$, to be either $Pr=1$ (with $\nu=\kappa=10^{-3}$) or $Pr=6$ (with $\nu=2.0\times10^{-3}$ and $\kappa=3.33\times10^{-4}$). Finally, $p$ is the pressure.
 
We define as usual the dimensionless Reynolds, Taylor Reynolds, and Froude numbers as 
\be 
Re = U_{rms}L_{int}/\nu ~ ~ , ~ ~ \ R_\lambda = \sqrt{\frac{5}{3}} \frac{U_{rms}^2}{\langle \boldsymbol{\omega}^2\rangle^{1/2} \nu} \ \ , \ \ Fr = U_{rms}/[L_{int} N] \ ,
\label{EQ:RE}
\ee
with $\vomega= \nabla \times {\bf u}$ the vorticity and $\lambda=\sqrt{5 U^2_{rms}/3 \langle \boldsymbol{\omega}^2\rangle}$ the Taylor scale.
 In Eq.~\eqref{EQ:RE} above, $L_{int}$ and $U_{rms}$  are the characteristic length-scale and velocity of the flow, with their ratio defining the turnover time $\tau_{NL}=L_{int}/U_{rms}$ \cite{marino_22}. In practice, $U_{rms}$ is the ({\it root mean squared}) velocity defined as $U_{rms} = \langle u^2 + v^2 + w^2 \rangle^{1/2}$, and $L_{int} = L_{tot}k_{min}/k_F^{(PP),(TG)}$.
 
An important derived parameter is the buoyancy Reynolds number $R_{IB}$ ~\cite{ivey_08} (see also \cite{davidson_13}), with
 $\varepsilon_V\equiv \nu \langle|\nabla {\bf u}|^2\rangle $ the global kinetic energy dissipation, brackets implying spatial averaging:
\be
\ R_{IB}= \varepsilon_V/[\nu N^2] \ . 
\label{EQ:RB}
\ee
We can also define the efficiency of kinetic energy dissipation $\beta$ when compared to a dimensional estimate $\varepsilon_D$, namely:
\be 
\beta\equiv \varepsilon_V/\varepsilon_D \ \ , \ \ \ \varepsilon_D=U_{rms}^3/L_{int} \ \ ; \ \ \ R_\mathcal{B} \equiv R_{IB} / \beta = \ ReFr^2  \ .\label{EQ:BETA}
\ee
The $R_{IB}$ parameter in equation (\ref{EQ:RB}) is sometimes used to denote the so--called interaction parameter \cite{gallon_25} (or the turbulent intensity \cite{pouquet_18}), while the buoyancy Reynolds number can also be found in the literature to be defined as 
$Re Fr^2$ \cite{feraco_18}. In this paper, we simply consider $R_{IB}$ as the buoyancy Reynolds number.

\begin{table*}[htbp]
\setlength\tabcolsep{5pt}
\begin{tabular}{|c|c|c|c|c|}
 \multicolumn{3}{l}{}\\ \hline
\rule{0pt}{4ex} \shortstack{\textbf{Series $\rightarrow$\ } \\ ~} & 
\shortstack{PP-Pr1 \\ \textbf{\#19 runs}} & 
\shortstack{PP-Pr6 \\ \textbf{\#17 runs}} & 
\shortstack{TG-Pr1 \\ \textbf{\#19 runs}} &
\shortstack{Passive Scalar \\ \textbf{\#5 runs}} \\
 \hline
 \rule{0pt}{3.5ex} $Pr$ & $1$ & $6$ & $1$ & $0.66$\\
 \hline
 \rule{0pt}{3.5ex} $Re$ & $2700\;\text{--}\;3400$ & $1300\;\text{--}\;1700$ & $5100\;\text{--}\;7400$ & $1400\;\text{--}\;9800$ \\
 \hline
 \rule{0pt}{3.5ex} $R_{\lambda}$ & $108\;\text{--}\;230$ & $72\;\text{--}\;139$ & $142\;\text{--}\;291$ & $78\;\text{--}\;210$ \\
 \hline
 \rule{0pt}{3.5ex} $Fr$ & \begin{tabular}{@{}c@{}}
$0.014\;\text{--}\;0.95$\\ (24, 515)*  \end{tabular}& 
$0.013\;\text{--}\;0.93$ & $0.013\;\text{--}\;1.04$ & $\infty$ \\
 \hline
 \rule{0pt}{3.5ex} $\beta$ & $0.11\;\text{--}\;0.46$ & $0.14\;\text{--}\;0.42$ & $0.14\;\text{--}\;0.42$ & $0.42$\\
 \hline
 \rule{0pt}{3.5ex} $R_{IB}$ & \begin{tabular}{@{}c@{}} $0.067\;\text{--}\;1132$\\ ($8.6\times10^5,\;\;4.0\times10^8$)*\end{tabular}& 
 $0.04\;\text{--}\;492$ & $0.23\;\text{--}\;2320$ & $\infty$ \\
 \hline
 \rule{0pt}{3.5ex} $R_\mathcal{B}$ & \begin{tabular}{@{}c@{}}
 $0.63\;\text{--}\;2712$\\ ($1.9\times10^6,\;\;8.6\times10^8$)*\end{tabular}& 
 $0.28\;\text{--}\;1280$ & $1.3\;\text{--}\;5485$ & $\infty$ \\
 \hline
\end{tabular}
\caption{Ranges of variation of characteristic parameters for the four series of runs analyzed herein, for a total of 60 runs:
Prandtl number $Pr$, Reynolds number $Re$, Taylor Reynolds number $R_\lambda$, Froude number $Fr$ (* runs with negligible stratification), dissipation efficiency $\beta$, 
and buoyancy Reynolds number $R_{IB}$ (as well as $R_\mathcal{B}$, see equations (\ref{EQ:RE}) -- (\ref{EQ:BETA})).
All parameters are temporally averaged over $5$ to $10$ turnover times, $\tau_{NL}$, after the peak of dissipation has been reached. Across the first three series for the stratified case, the runs all have a resolution of $512^3$ grid points and are forced at large scales; the runs labeled PP-Pr1 and PP-Pr6 use a random Pouquet-Patterson forcing, whereas TG-Pr1 employs a Taylor-Green forcing. The resolutions used for the passive scalar runs (last column) range from $128^3$ to $512^3$. }\label{Tab1}
\end{table*}

\noindent 
The buoyancy Reynolds number $R_{IB}$ can be seen as either a ratio of characteristic length scales, or as a ratio of characteristic time scales, 
{\it viz.} $R_{IB} = (\ell_{Oz}/\eta)^{4/3} = (\tau_\eta N)^{-2}$, 
where the typical time scale for buoyancy oscillations is $\sim 1/N$, $\tau_\eta$ is the time scale associated with the Kolmogorov dissipation length $\eta$, and $\ell_{Oz}$ is the Ozmidov scale; these scales are defined as usual as: 
 \be
 \eta=[\nu^3/\epsilon]^{1/4} \ \ , \ \ \ell_{Oz} = [\varepsilon_V/N^3]^{1/2} \ .  \label{eq:eta+Oz}
 \ee
 For a length-scale $\ell =\ell_{Oz}$, the buoyancy period $1/N$ and the turbulent turnover time $\tau_{NL}$ become comparable (assuming a classical small-scale Kolmogorov (1941) energy spectrum). 
For a review and a comparison with the related parameter $R_\mathcal{B} = Re Fr^2$, see also \cite{ivey_08}. 
The $R_\mathcal{B}$ and $R_{IB}$ parameters both measure the relative strength of buoyancy to dissipation and are commonly used to distinguish between wave-dominated and turbulence-dominated regimes \cite{smyth_19, pouquet_19p}.
 A transition from eddy-dominated to wave-dominated flows has also been recently observed at $R_{IB} \approx 1$ for both the motion of single particles~\cite{buaria_20} and for pairs of particles~\cite{gallon_24, gallon_25}.
 
We define as well, point-wise, the vertical flux of temperature fluctuations, or in short the buoyancy flux:
\be B_f= N w\theta \ . \label{BF} \ee
Finally, a measure of small-scale anisotropy in these flows can be obtained through the ratio of the dissipation due to vertical shearing of the horizontal wind to the total kinetic energy dissipation $\varepsilon_V$, namely \cite{brethouwer_07}
\be \pazocal{S}_a = \nu \langle(\partial_z u)^2+ (\partial_z v )^2 \rangle/ \varepsilon_V  \ .
\label{EQ:SHEAR} \ee
{In the case of homogeneous turbulent flows, standard estimates show that the averaged value $\langle \pazocal{S}_a \rangle = 4/15$. The latter is indeed recovered in the limit of very weak stratification, $N \rightarrow 0$.}

The numerical results concerning stratified flows analyzed herein were obtained by solving equations (\ref{momentum}) and (~\ref{scalar}) numerically using the Geophysical High-Order Suite for Turbulence (GHOST), a pseudo-spectral code that employs a hybrid parallelization combining MPI, OpenMP, and CUDA \cite{mininni_11h, rosenberg_20}. 
It allows for a variety of physical solvers, supports non-cubic geometries \cite{Sujovolsky,Alexakis_2024,Alexakis_2026}, 
and implements non-periodic boundary conditions in one (vertical) direction as well \cite{fontana_20}. 

The governing equations are integrated on isotropic grids of $512^3$ points and for $13.5\tau_{NL}$ to $24\tau_{NL}$. As stated earlier, the size of the periodic three-dimensional computational box is $L_{tot}=2\pi$. All in all, we analyzed several groups of simulations, divided as follows: 
19 (sharing parameters with Feraco et al. 2018, 2021 \cite{feraco_18,epl2021}, Marino et al. 2022 \cite{marino_22}, and Foldes 2025 \cite{foldes_25}) used the Pouquet–Patterson random forcing and $Pr = 1$, hereafter referred to as PP-Pr1; 17 more runs had the same random forcing but with $Pr = 6$, hereafter PP-Pr6; finally, 19 extra runs employed the Taylor–Green forcing with $Pr = 1$, hereafter TG-Pr1. 
{Additionally, we also solved the passive scalar equation, which can be viewed as a special case of Eqs.~(\ref{momentum}) and (\ref{scalar}), simply by 
writing $N=0$, as discussed in more detail in Section~\ref{subsec:weak_N}. Five runs with $N=0$ were performed on grids from $128^3$ to $512^3$ points. We picked an averaged scalar gradient $\mathbf{G}$ of unit length, parallel to the direction of one of the 3 coordinates. Further information on the code used in these cases can be found in~\cite{pumir_94}. A value of the Prandtl number of $Pr = 2/3$ was chosen, 
and the spatial resolution was maintained in such a way that $k_{max} \eta = 1.6$.}
The range of the main governing parameters for each series is reported in Table~\ref{Tab1}. With the exception of the passive scalar simulations, within each series, runs differ by the value of $N$, and for all of them we investigate the velocity field $\mathbf{u}$, temperature fluctuations 
$\theta$, and the buoyancy flux $B_f$ using high-order statistics. The kinematic viscosity and diffusivity were chosen such that in all simulations the Kolmogorov dissipation scale satisfies the condition $k_{max} \eta > 1$ at all times, where $k_{max}$ is the maximum numerically resolved wave number.

As a last point, note that we characterize as usual
the statistical distribution of a generic variable $\chi$ using dimensionless third- and fourth-order moments, skewness $S_\chi$ and kurtosis $K_\chi$, with a Gaussian distribution having zero skewness and a kurtosis of 3:
\be
S_\chi= \frac{\left<(\chi - \left<\chi\right>)^3\right>}{\left<(\chi - \left<\chi\right>)^2 \right>^{3/2}}, \,\,\,\,\,\,\,\,
K_\chi = \frac{\left<(\chi - \left<\chi\right>)^4 \right>}{\left<(\chi - \left<\chi\right>)^2 \right>^2}.
\label{EQ:SKKU}
\ee
Averages can be taken over volume, over horizontal planes, or over time.
Here, we combine volume and time averages, and consider most of the duration of the computations. 
When $K_\chi<3$, the Probability Density Functions (PDFs) have wings narrower than for a Gaussian. 
On the other hand, for values of $K_\chi$ higher than 3, the PDF tails are wider than for a Gaussian distribution. This means that the field $\chi$ develops events which deviate strongly from its mean value: the higher the kurtosis, the more extreme the events. 
As observed in F18, the components of the horizontal velocity field, $u$ and $v$, showed no signs of large-scale intermittency: their kurtosis for all the runs do not exceed 3 in the vast majority of the cases, with PDFs being almost Gaussian or sub-Gaussian. 
On the other hand, the vertical components of the velocity field (both Lagrangian and Eulerian) as well as the temperature field all develop, in some cases, extreme events in the flow (as in the flow with $Fr=0.076$, $N=8$ in F18 and M22, here run 9); these events are associated with values of the kurtosis higher than 3 and non-Gaussian PDFs \cite{feraco_18,pouquet_19p,epl2021, marino_22,reartes_26}. 
Herein, we want to extend the previous studies mentioned above, concentrating in particular on the buoyancy flux $B_f$, and on the relationship between extreme events in $B_f$, in $w$, and $\theta$. 

\section{Large-scale intermittency in stably stratified turbulence}  \label{S:INT}
\subsection{Overview} 
Stratified turbulence, stable or convective, is well-known to be intermittent, a phenomenon often measured in terms of non-Gaussian wings of probability density functions of relevant fields. A passive scalar 
advected by a turbulent velocity has strong intermittency in the small scales \cite{pumir_94}, as diagnosed by anomalous exponents of the two-point structure functions of its gradients. At 
large scales, intermittency in the planetary boundary layer for example, is measured in terms of one-point PDFs of thermal plumes which are seen to be responsible for the anisotropy of the flow \cite{chowdhuri_20}. Network metrics can allow for a quantification of the characteristics of strong events in the form of turbulent plumes, as shown in the context of a laboratory experiment of a turbulent boundary layer when analyzing turbulent transport \cite{iacobello_19,barulli_26}. Strong events in the nocturnal planetary boundary layer and in the atmosphere in general \cite{mahrt_89,sardeshmukh_15, fritts_17} have been observed as well as in the oceans \cite{klymak_06, dasaro_07, vanharen_16j}. In fact, numerous observational and numerical studies have been devoted to this problem (see \citet{caulfield_21} and references therein for a recent review).

\begin{figure*} 
\centering
\includegraphics[width=0.95\textwidth]{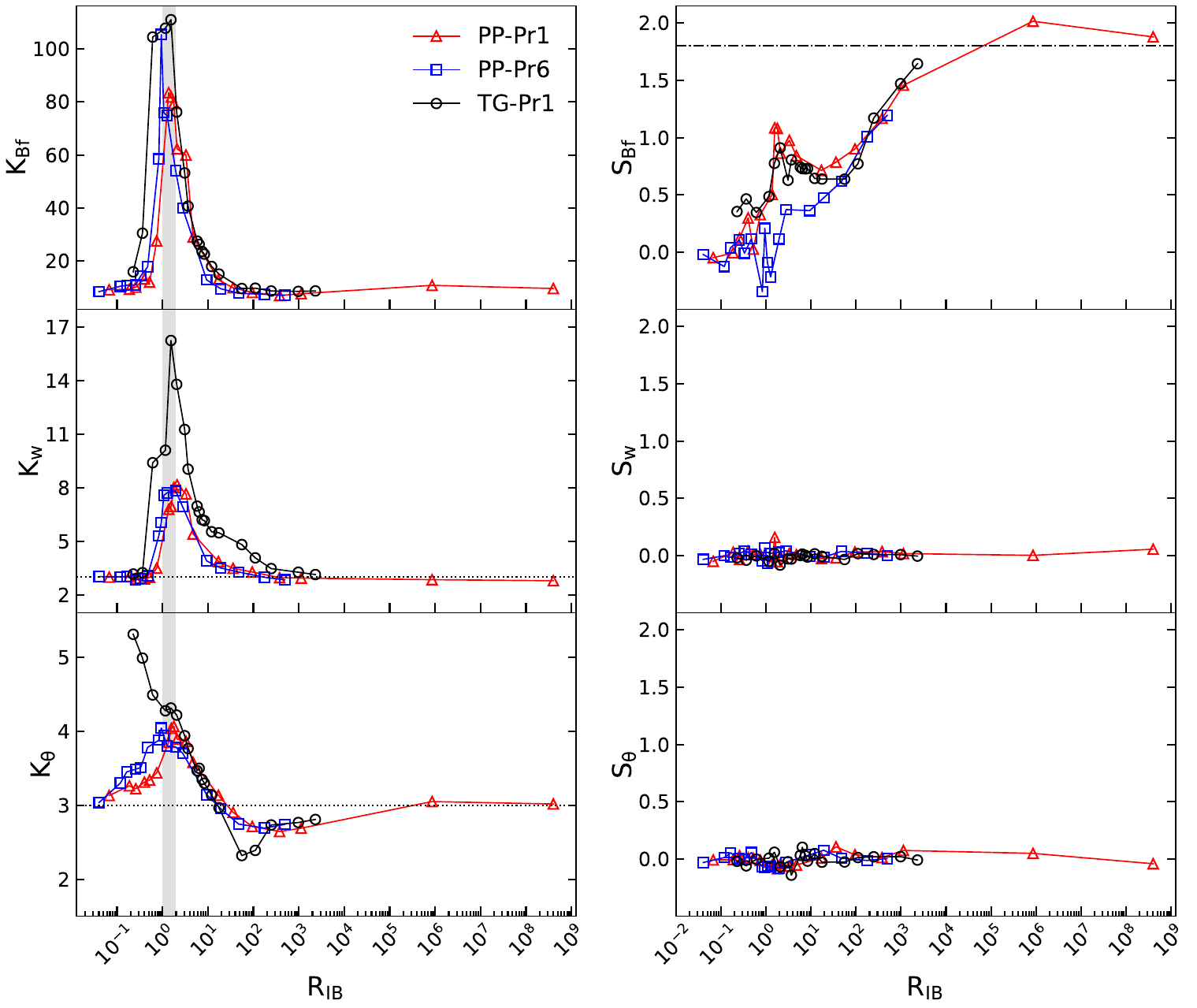}
\caption{
{\underline{{\it Top:}}} Kurtosis (left) and skewness (right) of the buoyancy flux $K_{B_f}$, $S_{B_f}$ as a function of buoyancy Reynolds number $R_{IB}$ for the three simulations series of this study (see Table~\ref{Tab1}); dot-dashed line in the top right panel indicates the value of the skewness of the vertical scalar flux, $S_{\rho w}\simeq 1.8$, estimated for the simulations with the passive scalar $\rho$, and horizontal dotted lines are for the Gaussian reference for the kurtosis which is $=3$.
{\underline{{\it Middle:}}} Kurtosis and skewness of the vertical velocity $K_w$, $S_w$ for the same runs.
{\underline{{\it Bottom:}}} Kurtosis and skewness of the temperature fluctuation $K_{\theta}$, $S_{\theta}$ for the same runs. 
Note the changes of scale between graphs for the kurtosis data. 
The narrower grey region $1<R_{IB} <2$ indicates where the maxima of the peaks for the three series are situated. 
}
\label{fig:1} \end{figure*} 

In the present context, a detailed numerical analysis of stratified turbulence was performed in the idealized geometry of periodic flows on cubic grids of mostly $512^3$ points \cite{feraco_18,marino_22}. It was found that for some values of the Froude number, the vertical velocity and the temperature fluctuations have non-Gaussian wings (see R14), and that in fact this phenomenon occurs in a narrow range of Froude numbers centered on $Fr\approx 0.076$ (or $N=8$, see F18 and M22).
These strong events take place in correspondence of overturning regions, for gradient Richardson number 
 close to 1/4 (and smaller), with thus a direct link to instabilities and dissipation \cite{sujovolsky_20, epl2021, marino_22}.
It has also been shown that large-scale intermittent vertical drafts may act as local energy injection mechanisms, also driving conversions between kinetic and potential energies \cite{foldes_25}. A simple model of intermittent behavior for fully developed turbulence to which the linear (wave) term was added, together with forcing and dissipation, could reproduce such a behavior, with the characteristic scale of the phenomenon likely associated with the Ozmidov scale $\ell_{Oz}$ (see R14 and \S \ref{S:3B} below). 

The kurtosis of the vertical velocity can take extreme values in that narrow range of Froude numbers, and in this regime of wave-eddy balance, a quasi-linear scaling of the mixing efficiency with the Froude number was reported as well \cite{feraco_18}. This parametric study was pursued for substantially longer times, of the order of $\approx 500 \tau_{NL}$ in M22, always performing direct numerical simulations of the Boussineq equations without resorting to sub-grid scale models or hyperviscosity. It was then possible to show that such a behavior takes place sporadically, followed by quiet episodes with quasi-Gaussian large-scale dynamics, a temporal intermittency well described by a reduced model introduced in M22. In time, the energy spectra in the small scales can change by several orders of magnitude between these two phases of the flow. One can show further that the large-scale strong drafts and the small-scale bursts typical of classical turbulence intermittency are connected, the evolution of the kurtosis of $w$ with Froude number being correlated with that of most of the field gradient components (\citet{epl2021}). 

\begin{figure*}
\centering
\includegraphics[width=0.93\textwidth]{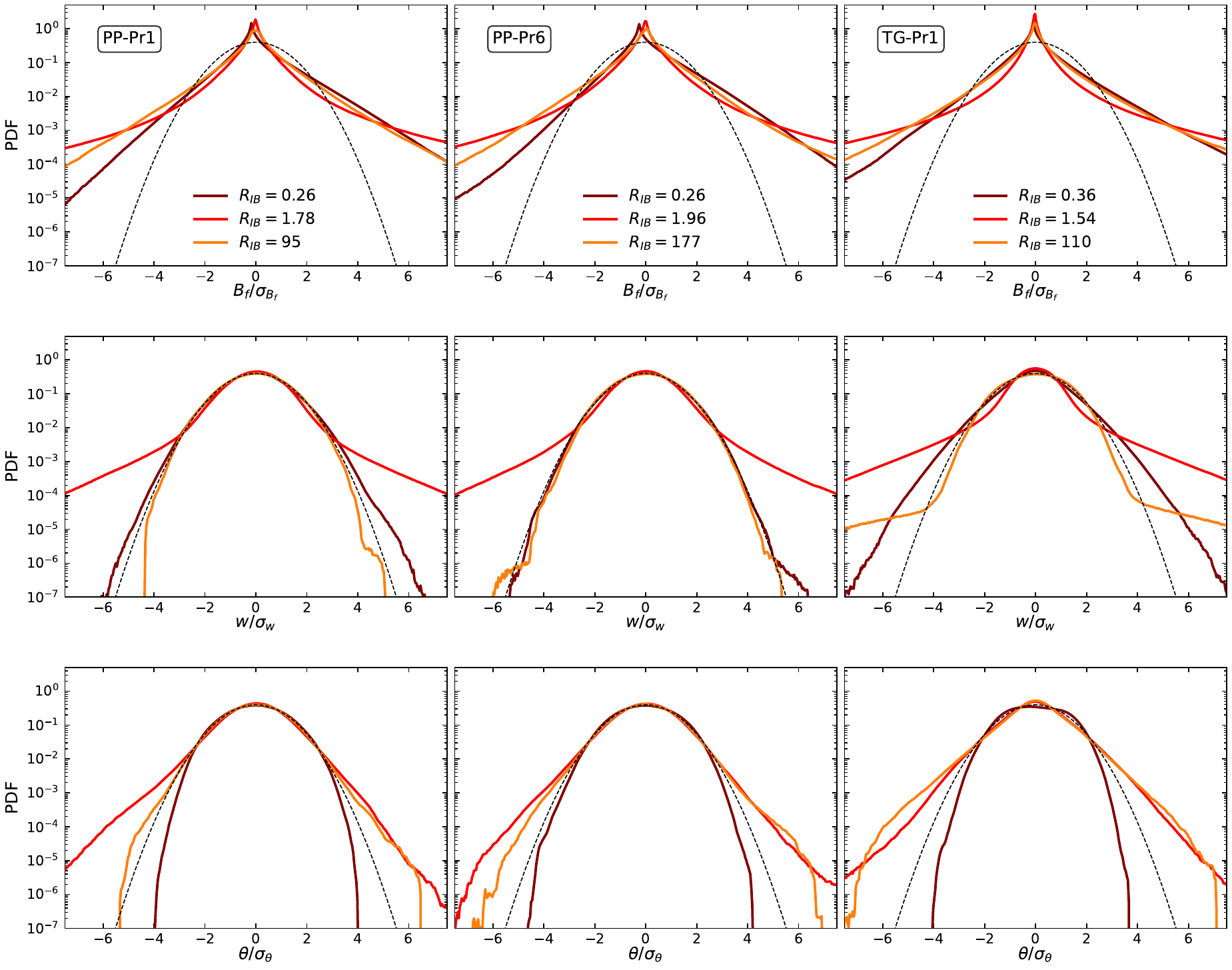} 
\caption{ 
Probability density functions (PDFs) of standardized buoyancy flux $B_f /\sigma_{B_f}$, vertical velocity $w/ \sigma_w$, and temperature fluctuation $\theta / \sigma_\theta$ (rows top to bottom) for three series datasets: PP-Pr1 (left column), PP-Pr6 (middle column), and TG-Pr1 (right column). 
Each variable is standardized by its global standard deviation prior to binning, and PDFs are accumulated over a statistically stationary time window. Colors denote the buoyancy Reynolds number $R_{IB}$: brown ($R_{IB} \approx 0.3,\; \mathcal{O}(10^{-1})$), red ($R_{IB} \approx 1.7,\;\mathcal{O}(1)$), and orange ($R_{IB} \approx 130,\;\mathcal{O}(10^2)$), corresponding to strongly stratified, transitional, and weakly stratified regimes, respectively. The dashed black curve is the standard Gaussian reference. Heavy tails relative to the Gaussian indicate large-scale intermittency.
}
\label{fig:3} 
\end{figure*} 

Furthermore, the sporadic and spatially-localized enhanced dissipation, both for the kinetic and the potential energy, is in agreement with measurements and simulations of oceanic dynamics; it shows in particular how a small volume of the ocean can lead to substantial energy dissipation even when the waves prevail at large scales when the Froude number is small \cite{klymak_08, pearson_18}. It was also found that where the strongest drafts develop, larger kinetic energy dissipation events are observed, and in fact these flows can be more intermittent than HIT in the sense that they require a smaller overall volume in order to achieve a given percentage of overall kinetic energy dissipation (see M22, Figure 5).

The physical mechanism behind these results can be understood qualitatively by a detailed mechanistic model based on the dynamics of the invariants of the velocity and temperature gradients \cite{sujovolsky_19, sujovolsky_20}. In this model, a closed set of equations for the Lagrangian evolution of velocity and temperature gradients is derived from the momentum and temperature equations under the Boussinesq approximation, and the occurrence of extreme events can be seen as the result of the system being at the brink of shear and convective instabilities (see more details below).
Finally, one can note that a similar phenomenon of strong large-scale intermittency 
occurs in rotating stratified flows, at least in a domain of parameters, pointing to the central role played by shear layers in turbulence \cite{barkley_16, shih_16, smyth_19, pouquet_19p, caulfield_21}.
A similar phenomenon has also been observed in space plasmas in the presence of a magnetic field \cite{marino_12}. 

\begin{figure*}
  \centering  
  \includegraphics[width=0.95\textwidth]{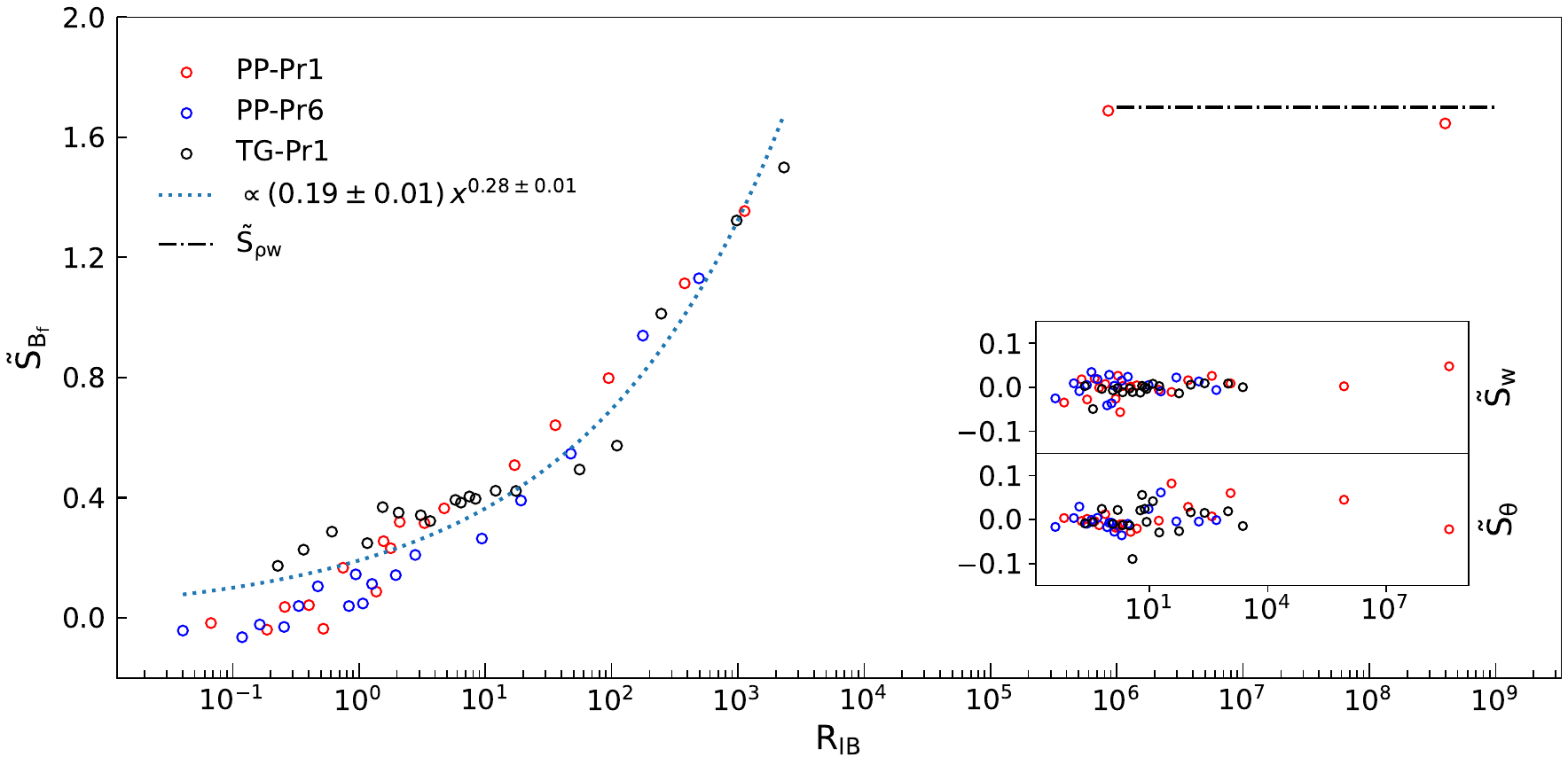}
  \caption{ 
Skewness of the filtered buoyancy flux $\tilde{S}_{B_f}$ in log-lin units for all stratified runs with values of $|w|$ and $|\theta|$ thresholded at their respective three sigma levels. The dashed line shows the fit  $\tilde{S}_{B_f} = (0.19\pm0.01) {R_{IB}}^{0.28\pm0.01}$. The horizontal  dot-dashed line at $R_{IB} \approx 10^{6}-10^8$ indicates the filtered passive scalar limit, $\tilde{S}_{\rho w}= 1.7$, estimated from our runs, which is close to the values of $\tilde{S}_{Bf}$ for the two most weakly stratified runs (marked * in Table~\ref{Tab1}).
The inset displays the corresponding thresholded skewness of $\tilde{S}_w$ (top) and $\tilde{S}_\theta$ (bottom), as a function of $R_{IB}$, for which no apparent scaling emerges.
  } 
  \label{fig:2} \end{figure*}

\subsection{Global trend and intermittent behavior of the buoyancy flux} \label{S:3B} 

We now concentrate on the intermittency of the buoyancy flux $B_f$, and contrast it with the intermittent behavior of the vertical velocity and temperature fluctuations previously documented in F18 and M22.

We computed the kurtosis of $w$, $\theta$, and $B_f$ for all the runs of the three simulation series of stratified flows analyzed herein, including for non-unitary $Pr$, and we find that they are characterized by non-Gaussian PDFs within overlapping ranges of values of the buoyancy Reynolds number. Indeed, the maxima of $K_w, K_{\theta}$ and $K_{B_f}$ all exceed $3$ in the range $1 < R_{IB} < 2$ (narrow grey shaded area in the left panels of Fig.~\ref{fig:1}). 
The only exception is represented by $K_{\theta}$ for the TG-Pr1 series, which keeps increasing as $R_{IB}$ decreases.
This could be related to the strong frontogenesis mechanism in TG flows \cite{sujovolsky_18, sujovolsky_20}.
Peaks of the kurtoses for the TG-Pr1 series occur instead at different values of Froude numbers than for the PP-Pr1 and PP-Pr6 series (not shown). In particular, the maxima of $K_{w,\theta,B_f}$ for the two series of runs driven with a Pouquet–Patterson random forcing occur at $0.05 < Fr < 0.08$, in agreement with F18, whereas they lie at $0.02 < Fr < 0.04$ for the series of runs employing the Taylor–Green forcing.
It is therefore worth emphasizing that $R_{IB}$ appears to be the parameter that best correlates with the emergence of large-scale intermittency in stratified turbulent flows, as observed in previous studies \cite{rorai_14,feraco_18,epl2021,marino_22}, in particular in F18. More importantly, the variation of the Prandtl number does not seem to affect the development of extreme vertical drafts, at least for the values considered in this study ($Pr = 1$ and $Pr=6$).

Thus, the first novel result of the present analysis is that not only do the PDFs of $w$ and $\theta$ develop strong non-Gaussian tails, but so do those of $B_f$ and at different Prandtl numbers, indicating that the buoyancy flux can be characterized by strong large-scale intermittency. This behavior shows clearly  when examining the PDFs of  $w$, $\theta$ and $B_f$, in runs characterized by $R_{IB} \approx \; \mathcal{O}(10^{-1}),  \;\mathcal{O}(1)$ and $\;\mathcal{O}(10^2)$, see Fig.~\ref{fig:3}.
These appear as spatially localized extreme events, scattered across the flow domain, where the kurtosis of $B_f$ attains values well above the Gaussian reference. Such behavior may also occur in geophysical flows, since both the atmosphere and the ocean can be characterized by local values of $R_{IB}$ close to unity, as defined here. In particular, $R_{IB}$ varies widely in the ocean, from $\mathcal{O}(1)$ in the thermocline to $\mathcal{O}(10^5)$ in the deep interior \cite{ivey_08}. 

We also show in Fig.~\ref{fig:1} (right panels) the skewness of the vertical component of the Eulerian velocity ($S_w$), of the temperature ($S_\theta$), and of the buoyancy flux ($S_{B_f}$), all given as functions of $R_{IB}$. Both $S_w$ and $S_\theta$ settle to values close to zero, consistent with distributions approaching Gaussianity at high $Fr$ and $R_{IB}$. 
This is not the case for the skewness of the buoyancy flux, which increases with $R_{IB}$ 
 (see Fig.~\ref{fig:1}, top right). The latter result may not be surprising, since $B_f$ is a quadratic quantity derived from these large-scale primitive variables. It is also the case because $B_f$ involves the correlation between $w$ and $\theta$. Even though both $w$ and $\theta$ individually approach Gaussian distributions ($S_w,~S_\theta \approx 0$), their joint probability distribution is non-Gaussian due to the physical coupling in stratified turbulence. 

To better characterize the trend followed by $S_{B_f}$, we recall that a kurtosis of the primitive fields significantly larger than 3 (Fig.~\ref{fig:1}) indicates the presence of extreme events, leading to wide tails of the PDFs of $w$, $\theta$, and $B_f$. The most intense vertical drafts and temperature bursts occupy only a marginal portion $p^V_{w,\theta}$ of the domain volume. We quantify this by selecting for each series, the run at the peak of $K_w$, where $p^V_{w,\theta}$ is expected to be largest. For PP-Pr1 run 9 ($R_{IB}=1.78$), the regions with $w/\sigma_w>3$ represent $p^V_w \approx 1.07\%$ of the fluid volume, while those with $\theta/\sigma_\theta>3$ occupy $p^V_\theta \approx 0.77\%$. For PP-Pr6 run 8 ($R_{IB}=0.95$), the corresponding fractions are $p^V_w \approx 0.64\%$ and $p^V_\theta \approx 0.76\%$. For TG-Pr1 run 5 ($R_{IB}=1.54$), $p^V_w \approx 1.87\%$ and $p^V_\theta \approx 0.93\%$.
Despite representing a very small fraction of the overall domain volume, large-scale intermittent drafts have been shown to significantly affect both small- and large-scale properties of the stratified flows supporting them, such as mixing and dissipation (as shown in F18 and M22), as well as particle transport \cite{reartes_26}.
Being the skewness sensitive to extreme values, to improve the identification of the scaling behavior of $B_f$, we therefore filter the values of  $|w|$ and $|\theta|$ exceeding three standard deviations, and denote by $\tilde{S}_{B_f}$ the skewness of the buoyancy flux computed from the filtered fields.
As expected, we found that those events still significantly contribute to $\tilde{S}_{B_f}$. This allows us to emphasize a well-defined scaling law for $\tilde{S}_{B_f}$,
which plausibly follows a power-law dependence, as shown in
Fig.~\ref{fig:2}, and characterizes the vast majority of the domain ($\sim 99\%$): 
$$
\tilde{S}_{B_f} = (0.19\pm0.01) {R_{IB}}^{0.28\pm0.01} \ \ \ 
$$
where the $\pm0.01$  reflect the standard error of the fit parameters.
This law holds up to $R_{IB} \lesssim 10^4$, beyond which $\tilde{S}_{B_f}$ saturates towards an asymptotic value of $\tilde{S}_{B_f} = 1.6$, close to that of filtered passive scalar limit $\tilde{S}_{\rho w} = 1.7$ estimated from our runs (dot-dashed line in Fig. \ref{fig:2}).
However, the skewness of both the filtered vertical velocity and the temperature remains roughly constant at each $R_{IB}$, with mean values close to zero $(\tilde{S}_w,~\tilde{S}_\theta \approx 0)$
(Fig. \ref{fig:2}, inset). 
High values of skewness of observable fields have been reported in the literature, for example in the free troposphere in the context of the reanalysis of climate data (27 years of ERA40 data taken daily) \cite{petoukhov_08}. The strong events generating them are associated with mid-latitude storms with strong baroclinic structures. It is further shown that temperatures are less intermittent than the vertical velocity, both in climate data reanalysis \cite{sardeshmukh_15}, as well as in our previous investigations (see Feraco et al.~\cite{epl2021}). 

The observed behavior can be captured by a model in the spirit of the work of Vieillefosse~\cite{vieillefosse_82, vieillefosse_84}, using the simplification that the pressure Hessian is isotropic. In the stratified case, caveats to this simplification \cite{Yang:24} are alleviated by the fact that waves can regularize the solutions \cite{sujovolsky_19}. Thus, to the original Vieillefosse model a wave term is added, as well as forcing and dissipation \cite{feraco_18}, together with coupling to temperature fluctuations. The sharp increase in the kurtosis reported here, as well as that reported for the Lagrangian particles vertical velocities in F18, are in fact a common feature of the intermittency in these flows as well as in the model. Let us write the model mentioned above for the strong intermittency of vertical velocity and temperature fluctuations as in R14 and F18, omitting dissipation and forcing:
\begin{eqnarray}
d_t \delta w &=& - \ \delta w^2/\ell_z \ \ \ - N \delta \theta \ , \label{modelW} \\
d_t \delta \theta &=& - \ \delta w \delta \theta/\ell_z +N \delta w \ ,  \label{modelT}
\end{eqnarray}
where $ \delta w$ and $\delta \theta$ are variations of the $w$ and $\theta$ fields on some unique and arbitrary scale $\ell_z$. 
This model is anisotropic in the sense that the velocity is restricted to its vertical component; $\ell_z$ is a characteristic vertical scale, and it was found in F18 that choosing $\ell_z$ proportional to the Ozmidov scale $\ell_{Oz}\equiv (\varepsilon_V/N^3)^{1/2}$ leads to the best fit of the model with the DNS data. Note that, under the assumption that a Kolmogorov isotropic scaling is established at scales smaller than $\ell_{Oz}$, and when the dissipation length $\eta$ is equal to $\ell_{Oz}$, then $R_{IB}=1$. Note also that an implicit assumption of the model is that horizontal components of the velocity and horizontal derivatives can be neglected because of the dominance of the vertical gradients and of the vertical component of the velocity field; this directly implies that the kinetic helicity 
$H_V= \langle {\bf u} \cdot \vomega \rangle$ is weak for these flows close to the peak of kurtosis and of strong large-scale intermittency, at least in domain regions perturbed by the extreme vertical drafts. This is in sharp contrast with small-scale intermittency of fully developed turbulence which resides in strong helical structures in the form of vortex filaments.
 
\begin{figure*} 
\includegraphics[width=16.cm]{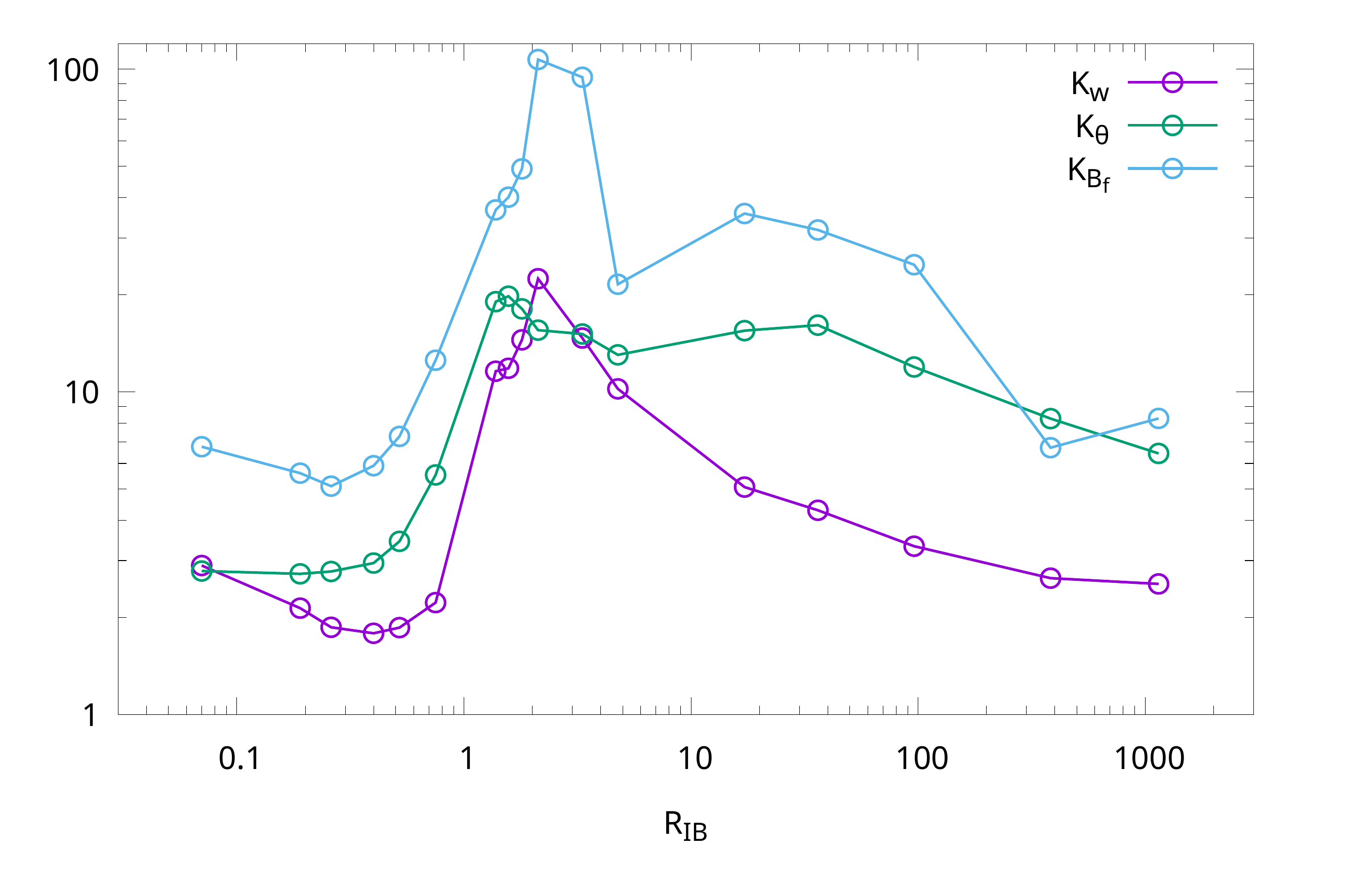}  
\caption{ 
Kurtosis of $w$, $\theta$ and $B_f$ obtained from the 1D model with input parameters taken from the runs of the PP–Pr1 series.}
\label{fig:4} 
\end{figure*}

In the framework of this system, it is straightforward to derive equations for the energy of the fluctuations at scale $\ell_z$, namely 
$\delta e_T= (\delta w^2+ \delta \theta^2)/2$, and for the (local in scale) buoyancy flux $\delta b_f= N \delta w \delta \theta$. They read:
\begin{eqnarray}
 d_t \delta e_T &=& -(2 \delta w/\ell_z) \delta e_T \ , \label{eq:model1}\\
 d_t \delta b_f &=& - (2 \delta w/\ell_z) \delta b_f - N^2 ( \delta \theta^2 - \delta w^2) \ .
 \label{eq:model2} \end{eqnarray}
Figure \ref{fig:4} shows the kurtosis of $w$, $\theta$ and $B_f$ obtained by integrating equations (\ref{eq:model1}) and (\ref{eq:model2}) after adding forcing and dissipation as in F18, using the same dimensionless parameters as those of the PP–Pr1 series (the two PP-Pr1 runs at very high $R_{IB}$, marked * in Table~\ref{Tab1}, are omitted).
We see that, for the buoyancy term, the equation becomes an energy-conservation statement in the Lagrangian sense (in the absence of forcing and dissipation as is the case here). On the other hand, the evolution of the buoyancy flux is altered by a differential between the potential energy and the vertical kinetic energy of the fluctuations on small scales of the order of $\ell_z$; this is similar of course in the primitive equations. Equipartition would make $\delta b_f$ evolve like $\delta e_T$, and it is in fact this lack of equipartition that makes $\delta b_f$ more susceptible to the local wave dynamics of energy exchanges.

Further note that $\ell_z$ governs the amplitude of the $\delta e_T,\ \delta b_f$ growth, and that the particular choice of $\ell_z=\ell_{Oz}$ mentioned above leads to a strong (quadratic) dependence of the outcome on the Froude number. Indeed, with $\ell_z=\ell_{Oz}=[\varepsilon_V/N^3]^{1/2}= L_0 [\beta Fr^3]^{1/2}$. Specifying also that we are in the intermediate regime of energy exchanges between nonlinear eddies and waves for which the efficiency of energy dissipation $\beta$ (see Eq.~\ref{EQ:BETA}) scales as $\beta \sim Fr$ as found for a parametric study of rotating stratified flows \cite{pouquet_18}, we obtain $\ell_z/L_0 \sim Fr^2$.
In that light, it would also be of interest to study, against the control parameter, the defect in equipartition between potential and kinetic (or simply vertical) energy, properly normalized. We already know that, in the transition regime between wave-dominated and turbulence-dominated flows, the ratio of potential to kinetic energy is rather constant and $approx 0.3$ \cite{pouquet_18}.
The drastic change in the kurtosis of $B_f$ observed in Fig. \ref{fig:1} can be explained by this extra source of variation just mentioned, {\it i.e.}, the kinetic-potential energy defect, but we also recall that $B_f$ is quadratic in the primitive variables, and does display a strong correlation between vertical velocity and temperature fluctuations.

As a final remark, we wish to point out that an improved model for the small-scale dynamics of stratified turbulence has been developed recently \cite{sujovolsky_19, sujovolsky_20}. It is based on the so-called PQR dynamics of the velocity gradient matrix \cite{chong_90} (P, Q and R being its invariants, with $P\equiv 0$ in the incompressible case). The model was extended to the stratified case (see M22) and it was shown that vertical gradients are prominent in this new model, a feature encompassed by construction in the system written in equations (\ref{modelW}) and (\ref{modelT}) since it is focused on the vertical velocity and vertical gradients introduced through 
$\ell_z$. The picture emerging from this refined analysis is one in which two slow invariant manifolds are attractive, one basically for gravity waves, and the other one for turbulence (and specifically, close to the threshold of the convective instability), where energy can be dissipated. Rapid transitions can occur between the two, with a relative residence time in one manifold with respect to the other one varying with Froude number. 
We also note that a number of studies have shown that stratified turbulence is in fact frequently close to an instability state, for example in the sense that the gradient Richardson number tends to be close (from below) to 1/4, {\it i.e.}, its classical critical value \cite{smyth_13, rosenberg_15, mashayek_17, smyth_19, pouquet_19p}. 

\begin{figure*} 
\centering
\includegraphics[width=1\textwidth]{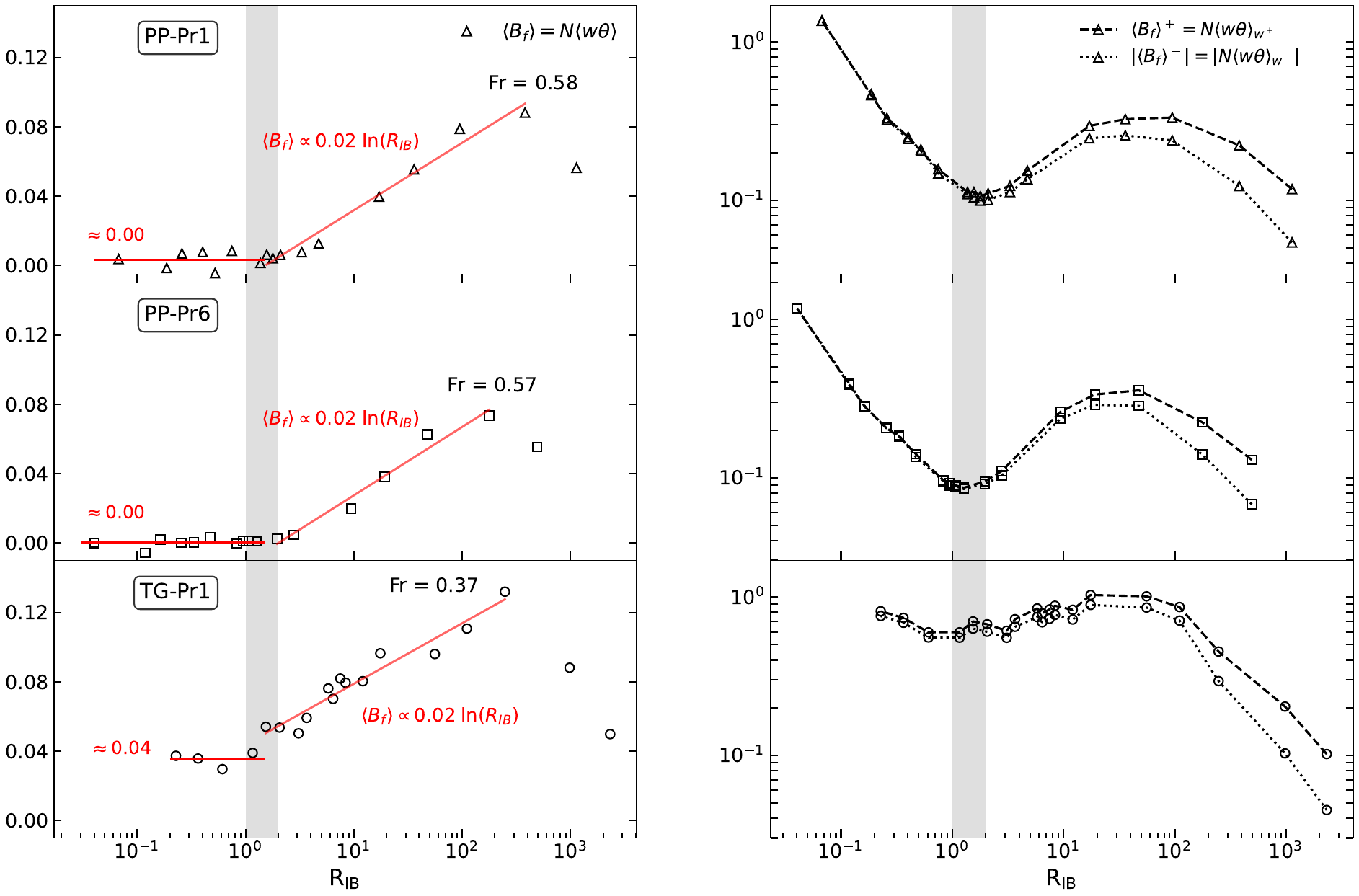} 
\caption{Scaling of the buoyancy flux as a function of the Buoyancy Reynolds. The three series are shown by rows : PP-Pr1 (top), PP-Pr6 (middle), TG-Pr1 (bottom). The left column shows the volume-averaged total $\langle B_f \rangle =N\langle w \theta \rangle$, while the right column shows its positive and negative contributions $\langle B_f \rangle^\pm$, conditioned on the relative signs of $w$ and $\theta$. The grey shaded region indicates the range containing the maxima of $K_{w, \theta, Bf}$ (see Fig.~\ref{fig:1}).
}
\label{fig:5}  \end{figure*} 

We now examine in more detail the structure of the buoyancy flux $B_f$ as a function of the buoyancy Reynolds number $R_{IB}$ for the three series of runs in Fig.~\ref{fig:5}. For clarity, the two  PP-Pr1 runs at very high $R_{IB}$ (marked * in Table~\ref{Tab1}) are omitted here to have comparable ranges of variation of $R_{IB}$ across the three series of runs. In particular, we give the volume averaged total $\langle B_f \rangle =N\langle w \theta \rangle$ in the left panels, and its positive and negative contributions $\langle B_f \rangle^\pm$ in the right panels, conditioned on whether $w$ and $\theta$ share the same or opposite signs, since $\langle B_f \rangle$ is positive when both have the same sign and negative otherwise.

The left panel of Fig.~\ref{fig:5} reveals three dynamically distinct regimes. 
For $R_{IB} \lesssim \mathcal{O}(1)$, $\langle B_f \rangle$ saturates to a offset value that may depend on the forcing, though is always very small: $\langle B_f \rangle \approx 0$ for both PP series, and $\langle B_f \rangle \approx 0.04$ for TG-Pr1, somehow reflecting the baseline level of $\langle B_f \rangle$ set by the large-scale forcing. The notably larger baseline for TG-Pr1 may be related to the substantially higher $U_{rms}$ in that series ($U^{\text{TG-Pr1}}_{rms} \approx 1.6 - 2.4$, compared to $1.0 - 1.4$ for the two PP series), though a definitive attribution requires further investigation. For $\mathcal{O}(1) \lesssim R_{IB} \lesssim \mathcal{O}(10^2)$, coinciding with the peak of the intermittency indicators $K_w, \ K_\theta$ and $K_{B_f}$, $\langle B_f \rangle$ begins to grow and follows a logarithmic scaling of the form:
$$
 \langle B_f \rangle \propto c\, \ln(R_{IB}) ,
$$
The slope $c \approx 0.02$ is consistent across all three series of runs, despite differences in forcing scheme and Prandtl number. Finally, for $R_{IB} \gtrsim \mathcal{O}(10^2)$, $\langle B_f \rangle$ reaches a maximum and subsequently decreases. The boundary of this third regime is identifed by values of the Froude number close to 0.5.
Recalling that $Fr = U_{rms} / [L_{int}N]$ can be interpreted as the ratio of the buoyancy time scale $\tau_b = 1/N$ to the nonlinear time scale $\tau_{NL} = L_{int}/U_{rms}$, values of $Fr\sim \mathcal{O}(0.5)$ indicate that the two time scales become comparable, beyond which turbulent overturning acts faster than stratification can restore it, and the net buoyancy flux declines.

The right panel of Fig.~\ref{fig:5} further shows that the range $1<R_{IB} <2$, where the maxima of $K_w, \ K_\theta$ and $K_{B_f}$ are found for all three series, is associated with an emerging imbalance between updrafts and downdrafts, with $\langle B_f \rangle ^\pm$ appearing to separate. This is clearer for the PP-Pr1 and PP-Pr6 runs. 
This imbalance grows and persists up to the highest value of $R_{IB}$ modeled.
In other words, the difference between the two $\pm$ sub-fluxes results in a comparatively small overall buoyancy flux at low values of $R_{IB}$, with net growth beginning only once the intermittency peaks near the transition into the second regime.

\begin{figure*} 
\centering
\includegraphics[width=0.85\textwidth]{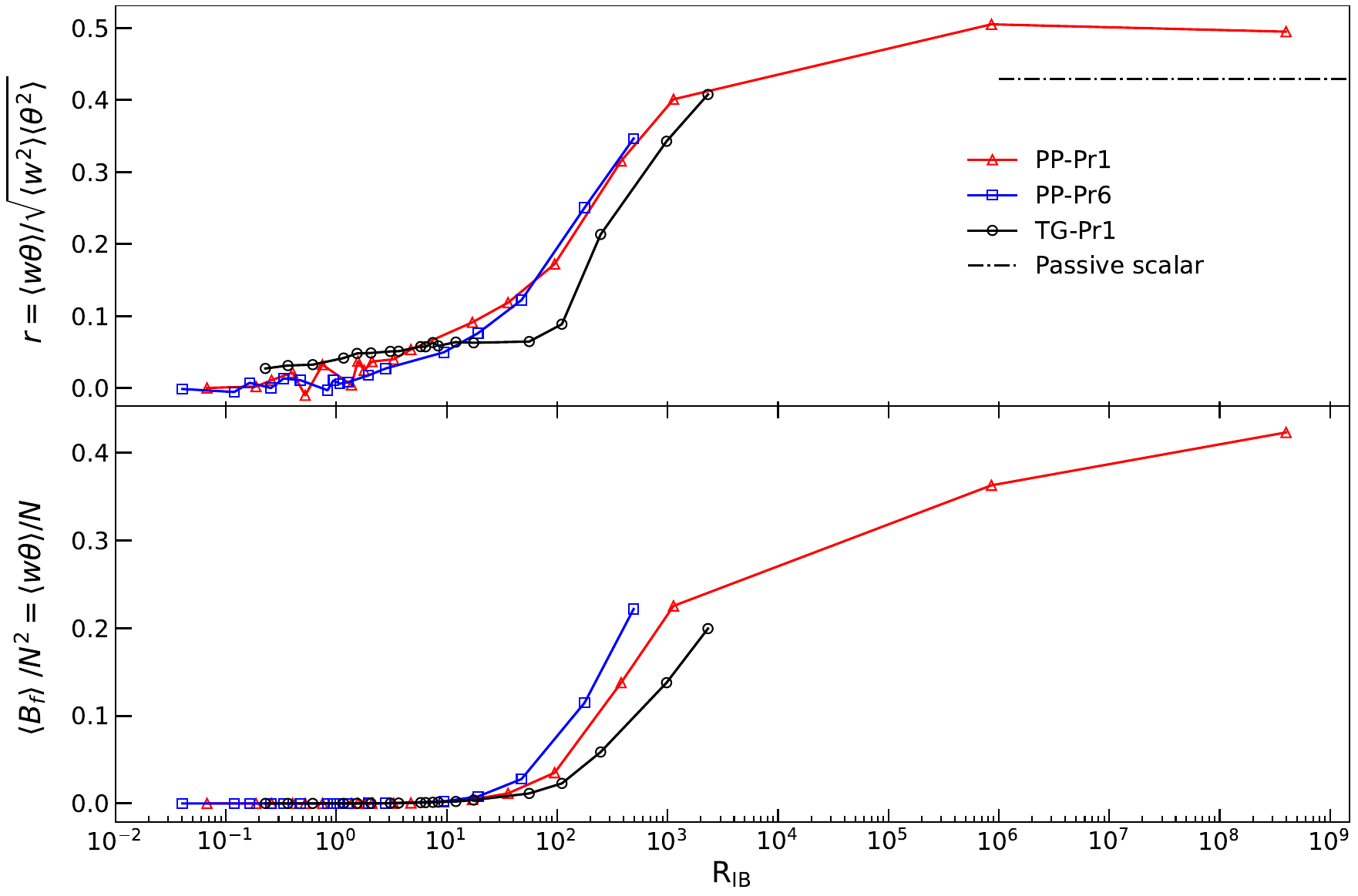} 
\caption{
{\underline{ {\it Top:}}} Pearson correlation coefficient $r_{w\theta}$ between the vertical velocity $w$ and temperature fluctuation $\theta$, as a function of $R_{IB}$ in log-lin coordinates. The dot-dashed horizontal line indicates the passive scalar reference value $r_{w\rho} = 0.44$, estimated from our passive scalar runs.
{\underline{ {\it Bottom:}}} The averaged total buoyancy flux $\langle B_f \rangle$ normalized by $N^2$, as a function of $R_{IB}$ in log-lin coordinates; note the saturation towards a constant value at high $R_{IB}$.
 }  \label{fig:6}  \end{figure*} 
 
Fig.~\ref{fig:6} shows the Pearson correlation coefficient $r_{w\theta}$ as a function of $R_{IB}$ for all three series of runs, now including the two extended PP-Pr1 runs (marked * in Table~\ref{Tab1}) which reach into the passive scalar limit. An abrupt increase in $r_{w\theta}$ is observed around $R_{IB}\approx 10$ for both PP sets and around $R_{IB}\approx 100$ for the TG run. In the passive scalar limit, $\theta$ is advected by the full three-dimensional velocity field $\mathbf{u}=(u,v,w)$, so there are also horizontal components that affect the scalar field, thus $r_{w\theta}< 1$. The observed saturation around 0.5 is consistent with our passive scalar runs, which yield $r_{w\rho}=0.44$, and also in agreement with other passive scalar DNS studies reporting similar values, $r_{w\rho} \approx 0.4-0.64$, depending on the Péclet and Schmidt numbers \citep{overholt1996, yeung2002}.
This behavior can be further understood by considering $N \to 0$ (very weak stratification) in the Boussinesq equations.
The correlation that builds up between the vertical component of velocity, $w$, and the rescaled scalar field $b = \theta/N$, leads to a finite production term for the scalar fluctuations, $\langle w b\rangle$. However, this only leads to a vanishingly small $\langle B_f \rangle = N^2 \langle w b \rangle$ when $R_{IB} \gg 1$. In quantitative terms, we observe that when $R_{IB} \to \infty$, $\langle B_f \rangle /N^2$ tends to a constant value of $0.42$. (see Fig. \ref{fig:6} bottom). This is further discussed below in \S.~\ref{subsec:weak_N}. 
\section{Limiting Behavior of the Boussinesq model for Extreme values of $N$} \label{S:LIMIT}
%
All simulations in this study are based on a parametric investigation of stratification, including the limit of very high values of the buoyancy Reynolds number, at which temperature fluctuations decouple from the dynamics and become passive. In our study, the selected stratification parameter, namely the Brunt–Väisälä frequency $N$, includes values of geophysical relevance. We now examine the extreme limits of the system, namely the two opposite limits $N \to \infty$ (strong stratification) and $N \to 0$ (weak stratification), both lying beyond realistic geophysical regimes.

\subsection{Strong stratification : $N\rightarrow \infty$}
In the strong stratification limit, the linear terms in Eqs.~\eqref{momentum} and \eqref{scalar} prevail, and the nonlinear terms become negligible, leading to the linearized system:
\begin{eqnarray}
\partial_t u  & = & - \partial_x p ~~ , ~~~ \partial_t v = - \partial_y p ~~, \nonumber \\
\partial_t w & = & - N\theta - \partial_z p ~~ , ~~~\partial_t \theta  =  Nw \ ,
\label{eq:waves}
\end{eqnarray}
which is the classical system describing waves in a stably stratified flow.
Note that in deriving these equations, we have omitted the dissipative terms, which act as damping. The solutions of Eq.~\eqref{eq:waves} are the classical waves solutions of the form $(u,v,w,\theta) = (u_0, v_0, w_0, \theta_0) \times \exp(i \mathbf{k} \cdot \mathbf{x} - \omega_0 t)$, where $\mathbf{k}$ and $\omega_0$ satisfy the standard dispersion relation $\omega_0^2 = N^2 k_z^2$~\cite{davidson_13}. The steady solutions of Eqs.~\eqref{eq:waves} correspond, on average, to a fixed amount of kinetic and potential energy. In the steady state, the term $N \langle w \theta \rangle$ describes an exchange between kinetic and potential energy, and therefore has to be $0$ on average: $\langle w \theta \rangle = 0$. Therefore, in the very strongly stratified case, $N \to \infty$, the solution is dominated by waves, which on the average maintain a fixed amount of kinetic and potential energy, therefore leading to a zero mean of the product of $w$ and $\theta$: $\langle \theta w \rangle / (\langle w^2 \rangle \langle \theta^2 \rangle)^{1/2} = 0$ (see the top panel of Fig. \ref{fig:6}).

\subsection{Weak stratification : $N\rightarrow 0$}
\label{subsec:weak_N}
In the limit of weak stratification ($N \to 0$), the Boussinesq equations reduce to the standard Navier-Stokes equations, the scalar field $\theta$ becoming effectively passive. This can be seen by defining a rescaled scalar field, 
\be
b \equiv \theta/N \ , \label{eq:b}
\ee
 so the gradient of $b$ becomes $\bf G = \hat{\mathbf{z}}$ ($| {\bf G} | = 1$. Substituting in Eqs.~(\ref{momentum},\ref{scalar}) leads to:
\begin{eqnarray} 
\partial_t {\bf u} + ({\bf u} \cdot \nabla) {\bf u} &=& - \nabla p - f{\bf \hat z} \times {\bf u}- N^2b {\bf \hat z} + \nu \nabla^2 {\bf u}, \label{eq:ns_mom}\\
\partial_t b + ({\bf u} \cdot \nabla) b &=& w + \kappa \nabla^2 b \ . \label{eq:PS} 
\end{eqnarray} 
In the limit $N \to 0$, the feedback of the scalar field $b$ on the momentum equation becomes negligible. In the passive ``potential energy'' problem, the term $\langle w \ b \rangle$ acts as a source term for the variance of the scalar fluctuations and is equal, in the statistically stationary state, to the scalar dissipation: $\langle w \ b \rangle = \kappa \langle (\nabla b)^2 \rangle = \epsilon_P$. In view of the scaling relation between $\theta$ and $b$, this production term is related to the buoyancy flux, $B_f$, defined by Eq.~\eqref{BF}, by $\langle B_f \rangle = N^2 \langle w \ b \rangle = N^2 \epsilon_P$. In the runs corresponding to the passive scalar problem, the scalar dissipation remains constant when viscosity decreases ($Re$ increases), which implies that $B_f \propto N^2$ when $N \to 0$. This leads to the
expectation that $\langle B_f \rangle \sim 1/R_{IB}$. 
Since $B_f$ and the production terms, $\langle w b \rangle$, are related by a numerical factor ($N^2$), their normalized higher moments, such as skewness of kurtosis, can be directly compared. 

\begin{figure*} 
\includegraphics[width=16.cm]{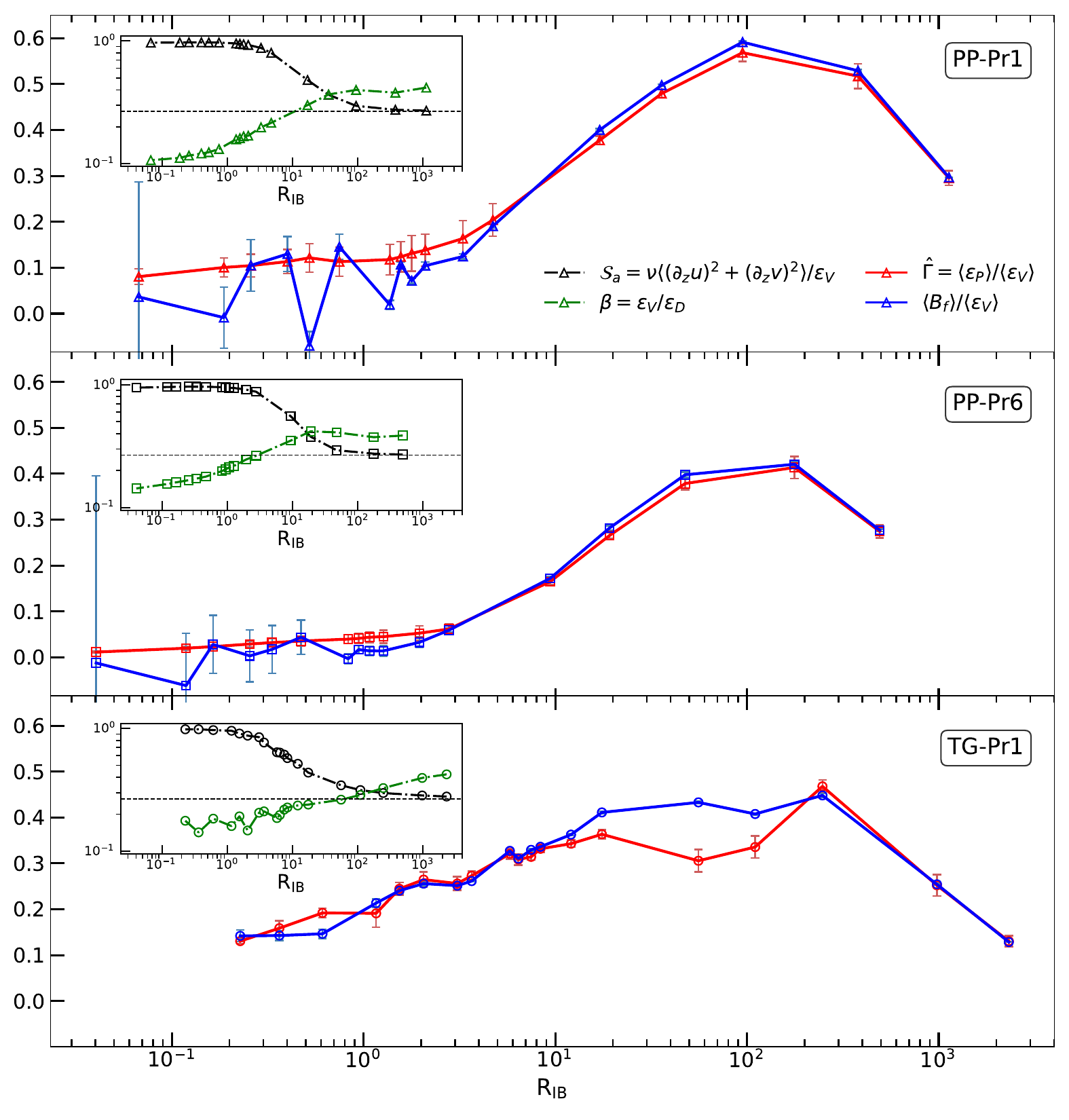} 
\caption{
Buoyancy flux $\langle B_f \rangle$ normalized by kinetic energy dissipation $\langle \varepsilon_V \rangle$ {\it i.e.} mixing efficiency (blue), and irreversible mixing efficiency $\hat \Gamma= \left< \varepsilon_P\right>/\left< \varepsilon_V \right>$ (red, Eq.~(\ref{EQ:GAM})) as a function of $R_{IB}$, for PP-Pr1 (top), PP-Pr6 (middle), and TG-Pr1 (bottom).
Insets show the normalized vertical shear $\pazocal{S}_a$ (black, Eq.~(\ref{EQ:SHEAR})) and kinetic dissipation efficiency $\beta = \varepsilon_V / \varepsilon_D$ (green) plotted on the same $R_{IB}$ axis; the horizontal dashed line marks the fully developed turbulence value $\pazocal{S}_a=0.267$. All quantities are spatio-temporally averaged over several eddy turn-over times.
} 
\label{fig:7}
\end{figure*} 

\begin{figure*} 
\includegraphics[width=14.cm]{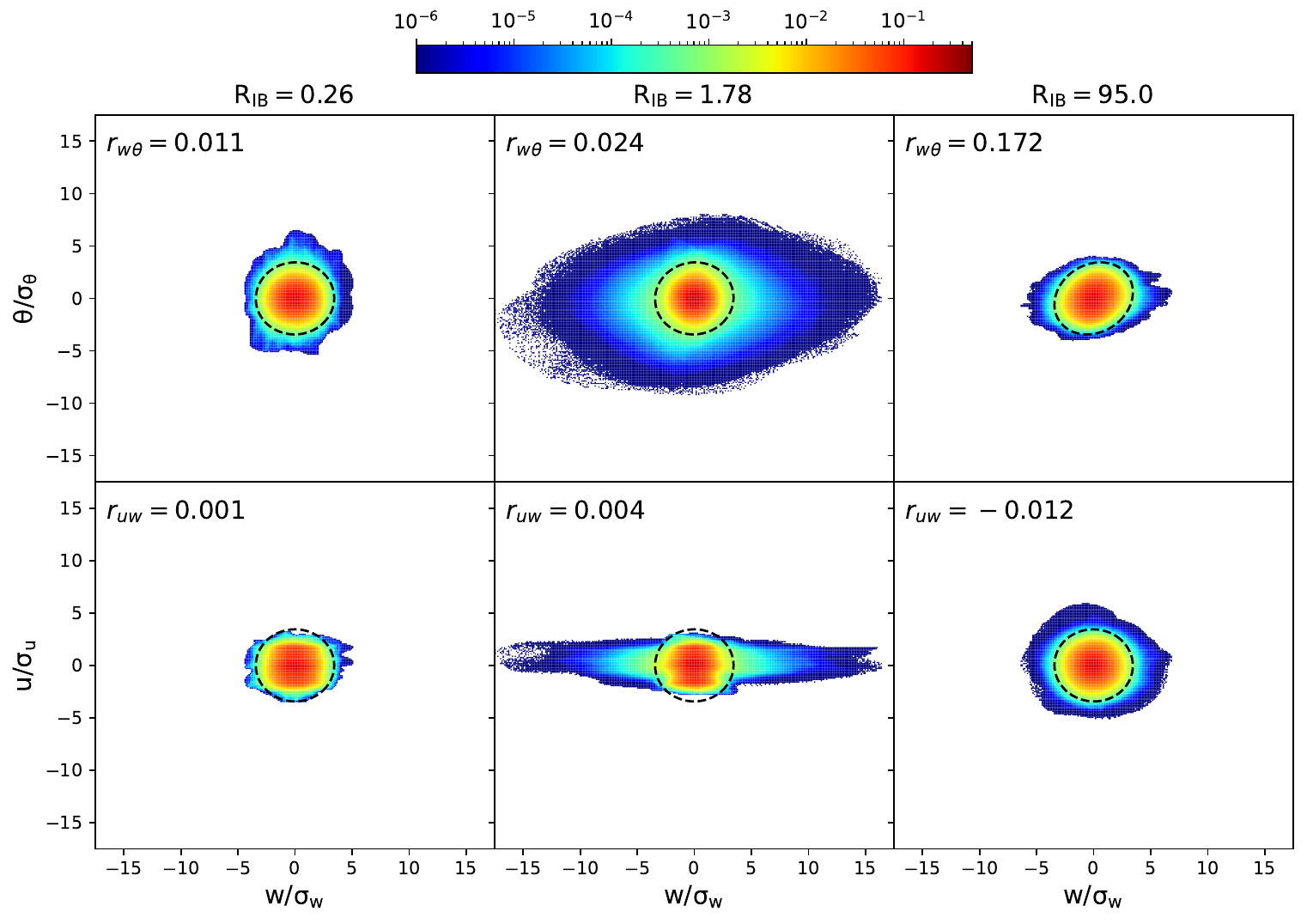} \vskip0.1truein
\caption{
{\underline{{\it Top}:}} Joint PDFs of temperature fluctuations $\theta$ {\it vs.} vertical velocity $w$, both normalized by their respective standard deviations, for three buoyancy Reynolds numbers: $R_{IB}=0.26$ (left, wave dominanted regime), $R_{IB}=\ 1.78$ (middle, peak of $K_w$) and $R_{IB}=95$ (right, turbulence-dominated regime).
{\underline{{\it Bottom}:}} Joint PDFs of the horizontal velocity, $u$, {\it vs.} the vertical velocity $w$, normalized by their respective standard deviations and for the same runs. The black dashed ellipses in each panel indicate the $3\sigma$ level of a 2D Gaussian distribution. The Pearson correlation coefficient $r$ between the two fields is indicated in each panel. All data are from the PP-Pr1 series.
}
\label{fig:8} \end{figure*} 

\begin{figure*} 
\includegraphics[width=0.45\textwidth]{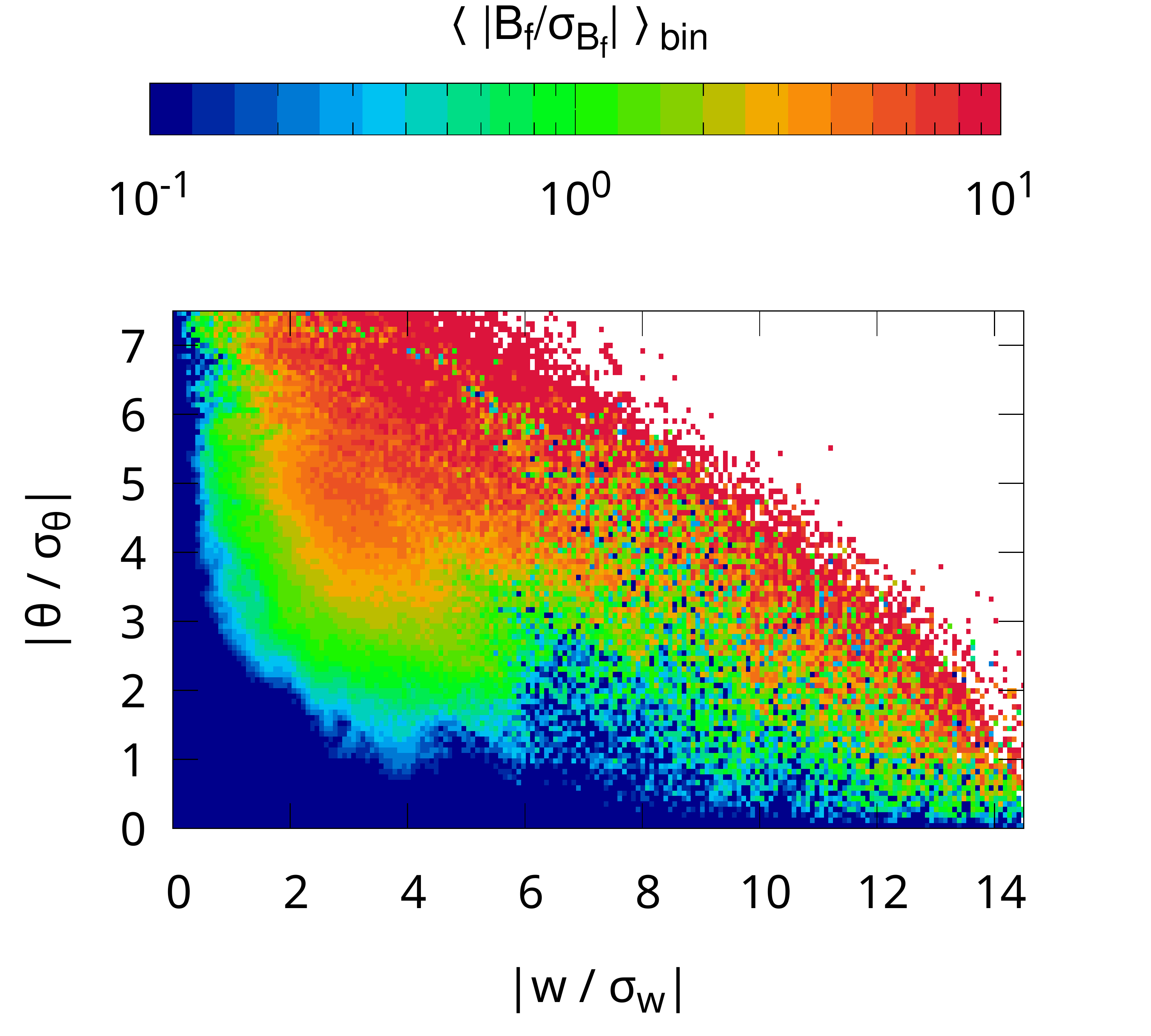}
\includegraphics[width=0.45\textwidth]{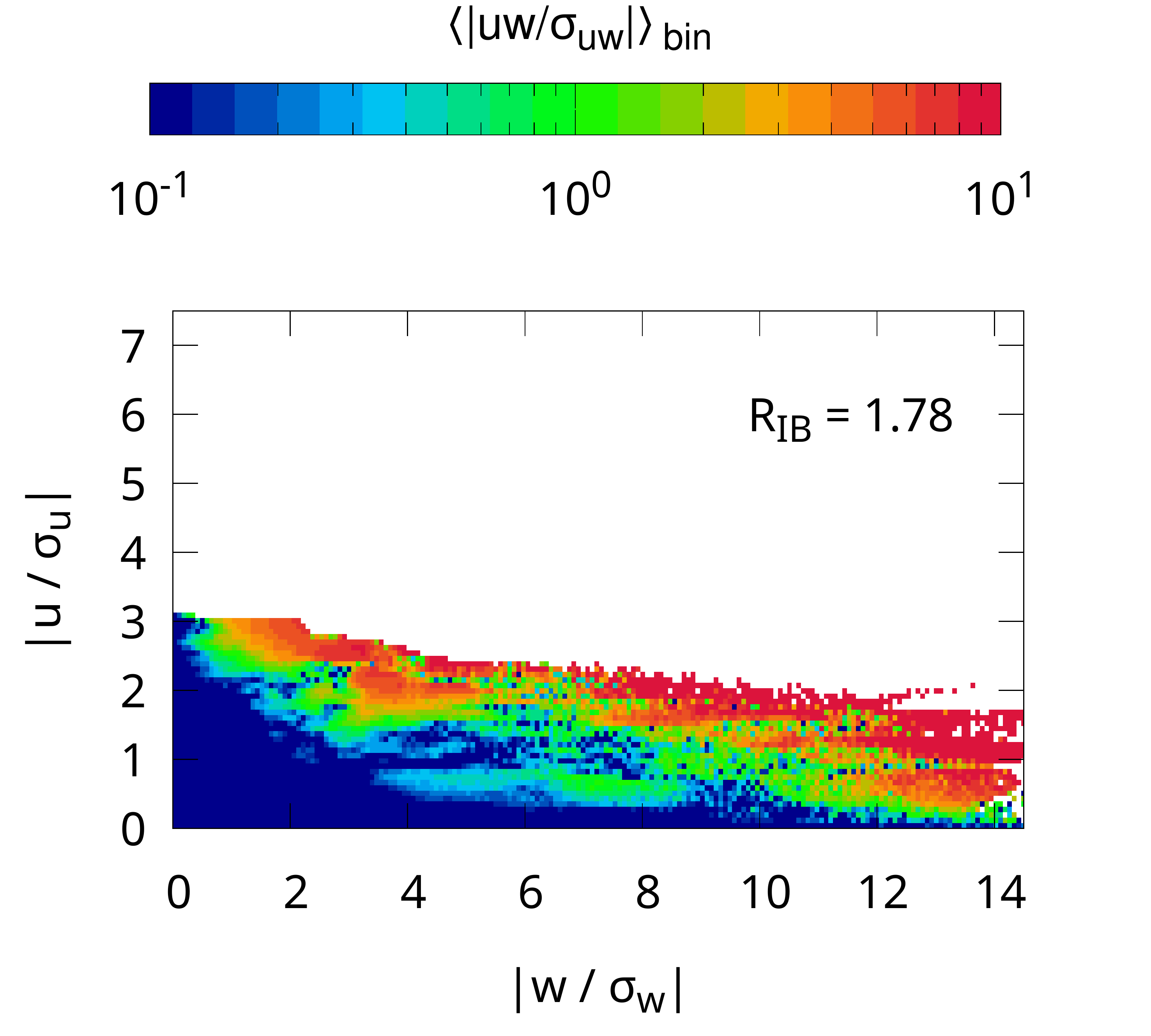}
\vskip0.1truein
\caption{
{\underline{{\it Left}:}} 2D map of the bin-conditioned mean $\langle |B_f / \sigma_{B_f}| \rangle_\text{bin}$, as a function of the absolute standardized vertical velocity $|w/\sigma_w|$ and absolute standardized temperature $|\theta/\sigma_\theta|$ (vertical axis).
{\underline{{\it Right}:}} 2D map of the bin-conditioned mean $\langle |uw / \sigma_{uw}| \rangle_\text{bin}$, as a function of $|w/\sigma_w|$ (horizontal axis) and $|u/\sigma_u|$ (vertical axis). Both panels show PP-Pr1 run 9 ($R_{IB}=1.78$, at the peak of $K_w$), with quantities accumulated over the same averaging window used throughout the analysis.
}
\label{fig:9} \end{figure*} 

\section{The link with anomalous mixing and dissipation} \label{S:DISS}
\subsection{Dissipation efficiency in stratified flows} 
We now analyze how stratification affects the mixing properties of the flow. The strong vertical drafts certainly provide a local contribution to mixing, but 
their quantitative importance for the overall
dynamics needs to be assessed.
In the deep ocean, mixing efficiency is found to vary by up to three orders of magnitude in several regions \cite{ijichi_18}, for buoyancy Reynolds numbers varying from 1 to $5000$ \cite{ivey_08}. This strong variations of mixing efficiency may well reflect the strong localization of turbulent hot spots where the dissipation occurs for high Reynolds number flows. 

The buoyancy flux $B_f$ is commonly used to characterize mixing in stratified turbulence and to evaluate the mixing efficiency defined here as 
$\langle B_f \rangle /\langle \varepsilon_V \rangle$. There are other possible definitions; for example, in F18, we looked at the mixing efficiency considering the available potential energy which can be converted into background potential energy, quantified by the rate of potential energy dissipation 
$\varepsilon_P= \kappa \langle\lvert \nabla \theta \rvert^2\rangle$, in order to evaluate the irreversible mixing efficiency defined now as (see also for example \cite{venayagamoorthy_16, pouquet_18}):
 \be
 \hat \Gamma =\langle \varepsilon_P \rangle / \langle \varepsilon_V \rangle\ .
  \label{EQ:GAM} \ee

We have already noted that a drastic change in the behavior of the high-order moment statistics of $w$, $\theta$, and $B_f$ occurs across the range $1<R_{IB}<2$.
This transition is also reflected in many of the statistics analyzed for the flows considered in this study. 
Fig.~\ref{fig:7} shows the averaged total buoyancy flux, normalized by $\langle \varepsilon_V \rangle$ (blue), defining the mixing efficiency, and the irreversible mixing efficiency $\hat \Gamma$ (red), both as a function of $R_{IB}$ for the three series of runs. The insets report the normalized vertical shear $\pazocal{S}_a$ (black) defined by Eq.~(\ref{EQ:SHEAR}), as well as the efficiency of kinetic energy dissipation $\beta = \varepsilon_V/\varepsilon_D$ (green).

For all quantities shown in Fig. \ref{fig:7}, three regimes can be distinguished: a low-$R_{IB}$ regime, where the normalized fluxes are quasi-constant (at a level that may depend on the Reynolds number \cite{pouquet_18} in this low-buoyancy Reynolds regime); a region of rapid increase around $1<R_{IB}<2$; and, at higher values of $R_{IB}$, either a plateau or a decrease. 
It is found in \citep{zhou_24} that the irreversible mixing efficiency (also called flux coefficient) is rather constant, with $\hat \Gamma \approx 0.47$
; our simulations are consistent with this value in the intermediate regime with $R_{IB}\approx 10^2$, for the TG-Pr1 showing this plateau over a wider range of $R_{IB}$.
 
In F18, $\hat \Gamma$ was presented as a function of the Froude number and was found to vary linearly over an intermediate range of Fr (see Fig.~5 of F18). In Fig.~\ref{fig:7}, $\hat \Gamma$ is plotted as a function of $R_{IB}$; since $R_{IB}=\beta Re Fr^2$, the linear dependence of $\hat \Gamma$ on Fr translates into power-law dependence $\hat \Gamma \propto {R_{IB}}^m$, with $m \approx 0.5$ expected when $\beta Re$ is constant. 
However, $\beta$ and $Re$ vary across the runs, so we estimate $m$ by a log-log linear regression of $\hat{\Gamma}$ against $R_{IB}$ over the intermediate range, yielding exponents in the range $0.12 \le m \le 0.51$ for the three series.

The consistent logarithmic scaling found for $\langle B_f \rangle$ across the three series (Fig.~\ref{fig:5}) does not seem to extend to $\hat{\Gamma}$.
On the other hand, $\hat{\Gamma}$ and $\langle B_f \rangle /\langle \epsilon_V \rangle$ show almost exact concordance across the runs, in particular for runs at and beyond the peak of $K_w$, confirming that the variation of $\hat{\Gamma}$ with $R_{IB}$ is largely governed by $\langle \varepsilon_V \rangle$. The only notable exception is at low $R_{IB}$, where $\hat{\Gamma}$ seems to better converge to a saturation value, within errors while $\langle B_f \rangle /\langle \epsilon_V \rangle$ fluctuates more.


The constancy observed for $\langle B_f \rangle /\langle \epsilon_V \rangle$ and $\hat \Gamma$ in the same range of low $R_{IB}$ likely indicates that shear layers are the dominant small-scale structures in this regime. One could alternatively say that the shear generates small-scales which dissipate the total energy, and the flow returns to a state dominated by waves, thus creating an energetic cycle. 
In the limit $N \rightarrow 0$, we are back to passive scalar dynamics with a mean gradient (neglecting the coupling term in Eq.(1)). But there are correlations between $\theta$ and $\mathbf{u}$, the flux disappearing because $\theta \propto N$, as explained in Section~\ref{subsec:weak_N}. This explains the unavoidable decrease of the mixing efficiency at high $R_{IB}$: as turbulence prevails over gravity waves, the scalar field becomes progressively more decoupled from the velocity field.


The behavior of the normalized vertical shear $\pazocal{S}_a$ (black, inset of Fig. \ref{fig:7}) was already noted before \cite{brethouwer_07}. In the case of the study of these authors, the forcing is purely two-dimensional, so that vertical fluctuations can develop under the coupling of the turbulence and the gravity waves. In their study, $\pazocal{S}_a \approx 1$ for $R_{IB} < 1$, reflecting dissipation dominated by vertical shearing of large-scale motions, while for $R_{IB} > 1$, $\pazocal{S}_a$ decreases and approaches the isotropic turbulence value $\pazocal{S}_a = 0.267$.
Our results show the same qualitative behavior and approach the same asymptotic value. The agreement between our curves and those of \cite{brethouwer_07} is consistent up to the maximum buoyancy Reynolds number of their study ($R_{IB} \approx 15.6$, in their Fig.~10). The inflection point marking the transition from the wave-dominated to the turbulence-dominated regime is located at $R_{IB} \approx 3$ in \cite{brethouwer_07}, whereas in our three series of runs it falls in the range $5.8\lesssim R_{IB} \lesssim 17$, consistent even with the different forcing schemes were employed.


Finally, the results for PP-Pr1 and PP-Pr6 appear nearly identical in shape, differing only in magnitude, suggesting that the Prandtl number does not play a pre-eminent role for the range of values considered here. This is consistent with the fact that increasing $Pr$ decreases $\varepsilon_P$ and conversely increase $\varepsilon_V$ \cite{riley2023}. It is also noted in \cite{riley2023} that large-scale structures are not obviously affected by changes in $Pr$ other than the fact that they lose energy at differing rates depending on $Pr$.

\subsection{Peak of kurtosis and large temperature and vertical drafts }   

The analysis that follows focuses on the PP-Pr1 series, whose outcome is analogous to those of PP-Pr6 and TG-Pr1. In Fig. \ref{fig:8} we show two-dimensional joint PDFs of vertical velocity $w$ (horizontal axis), against temperature fluctuations $\theta$ (top), and horizontal velocity $u$ (bottom), all normalized by their respective standard deviations $\sigma_u$, $\sigma_w$, or $\sigma_\theta$). The plots are shown for three buoyancy Reynolds numbers, namely $R_{IB}=0.26$ (run 3, left, for the wave-dominated regime), $R_{IB}=1.78$ (run 9, middle, at peak of intermittency), and $R_{IB}=95$ (run 15, right, in the stronger turbulence regime). All quantities are computed over the simulation domain at each available time output of the simulation, over the same time interval considered for the previous analyses.

The symmetries of the problem impose isotropy in the horizontal plane, so that there is no preferred direction for the momentum transport and $\langle u w \rangle = 0$ on average. On the other hand, the buoyancy flux is in general nonzero due to the coupling between vertical velocity and temperature fluctuation, with $\langle w \theta \rangle = r_{w\theta} \sigma_w \sigma_\theta$, where $r_{w\theta}$ being the Pearson correlation coefficient.
The dashed ellipses in Fig.~\ref{fig:8} show the 99.7\% ($3\sigma$) isoprobability contour fit of the bi-variate Gaussian distribution, determined by $\sigma_w$, $\sigma_\theta$ (or $\sigma_u$) and $r$. Values of $r$ are close to zero for the two lowest $R_{IB}$ and increase to $r_{w\theta} \approx 0.172$ at $R_{IB} = 96$, consistent with the progressive increase of the $w$-$\theta$ correlation as stratification weakens (see Fig. \ref{fig:6} top). Nevertheless, we observe that the distributions differ, depending on the values of $R_{IB}$ (or equivalently, $Fr$): for $R_{IB}=0.26$ (run 3, left) and $R_{IB}=95$ (run 15, right), the observed values of both $w$ and $\theta$ do not go beyond $7\sigma$, with run 3 (left) showing slightly stronger fluctuations in $\theta$ than in $w$. By contrast, $R_{IB}=1.78$ (run 9, middle) at the peak of $K_w$ exhibits the largest excursions, with $w$ reaching up to $15\sigma_w$ and $\theta$ up to $7\sigma_\theta$, reflecting the strong large-scale intermittency at this value of $R_{IB}$.

On the other hand, the horizontal velocity remains limited, except at the highest buoyancy Reynolds number when turbulence and isotropization both begin to develop. Indeed, the overall distribution for the highest $R_{IB}$ is globally circular (right panels), reflecting a progressive return to isotropy although this isotropization of the flow is slow to develop, in particular for the vorticity \cite{pouquet_19p}. 
Indeed, it is known that the return to isotropy in turbulent shear flows, as quantified by the two-point correlation function for the vorticity, 
$\langle \omega_i \omega_j \rangle$, and for large Taylor Reynolds numbers ($R_\lambda \to \infty$) is relatively slow, 
$\sim R_\lambda^{-1/2}$~\cite{Lumley67}. Furthermore, in our case, we notice that the flow is of almost zero divergence in the horizontal, due to its partial decorrelation with vertical dynamics at low Froude number and that, on the other hand, vertical gradients in stratified flows are high: the highest values of the velocity flux are obtained for the highest buoyancy Reynolds number run.
We also note that the core (low values) remain the same for the three runs, but what characterizes the peak of kurtosis in $w$ is the extension in accessible values of $\theta$, and even more so for $w$. In other words, these general arguments give a slow return of the vorticity towards isotropy as a function of $R_\lambda$, already observed for stratified rotating turbulence for example in \cite{pouquet_19p}.
The argument is perturbative in the following sense: the correlation function --- {\it e.g.} the spectrum --- is the sum of the isotropic contribution, plus a correction linear in the shear. One then has, as a dimensionless perturbation parameter, $S/\langle \omega^2 \rangle^{1/2}$. Writing now
$\langle \omega^2 \rangle \sim (U/L)^2 Re \sim R_\lambda^{1/2}$ leads to the desired result. 

 One clearly sees the greater extension of the temperature fluctuations compared to the horizontal velocity ones for the simulation $R_{IB}\approx 1.78$ (Fig.\ref{fig:8}, mid panels). 
Finally, we observe in the right panel a slight decline in $u$ as the (normalized and unsigned) vertical velocity $w$ increases; however, the largest momentum fluxes are obtained over a wide range of $w$ values, starting at $|w/\sigma_w| \gtrapprox 1$, similarly to the temperature case. For both the temperature and the horizontal velocity, high vertical velocities are obtained at the expense of low temperature fluctuations, and conversely the large temperature excursions are for order-unity vertical velocity. This clearly shows an energy exchange between these two variables, as in fact modeled by the system given in Eqs.~(\ref{eq:model1}) and (\ref{eq:model2}). 
Also note the low (red) pocket in temperature fluctuations $|\theta/\sigma_\theta|$ of high buoyancy fluxes showing the essential role of the coupling to the temperature fluctuations in obtaining the strong intermittency in run 9.

In order to further investigate the dependence of buoyancy and momentum fluxes on the primitive fields, Fig. \ref{fig:9} shows the bin-conditioned means $\langle |B_f / \sigma_{B_f}| \rangle_\text{bin}$ and $\langle |uw / \sigma_{uw}| \rangle_\text{bin}$, accumulated over the entire run with $R_{IB} = 1.78$ (run 9, corresponding to $Fr=0.076$ in F18 and M22), as a function of the standardized vertical velocity $|w/\sigma_w|$ and temperature $|\theta / \sigma_\theta|$ (left), or $|w/\sigma_w|$ and horizontal velocity $|u/\sigma_u|$ (right), respectively. These visualizations show that while the largest values of the buoyancy flux may occur due to bursts of either temperature or vertical velocity, instead large values of the momentum flux are driven by vertical velocity augmentations only.


\begin{figure*}
 \includegraphics[width=8.1cm]{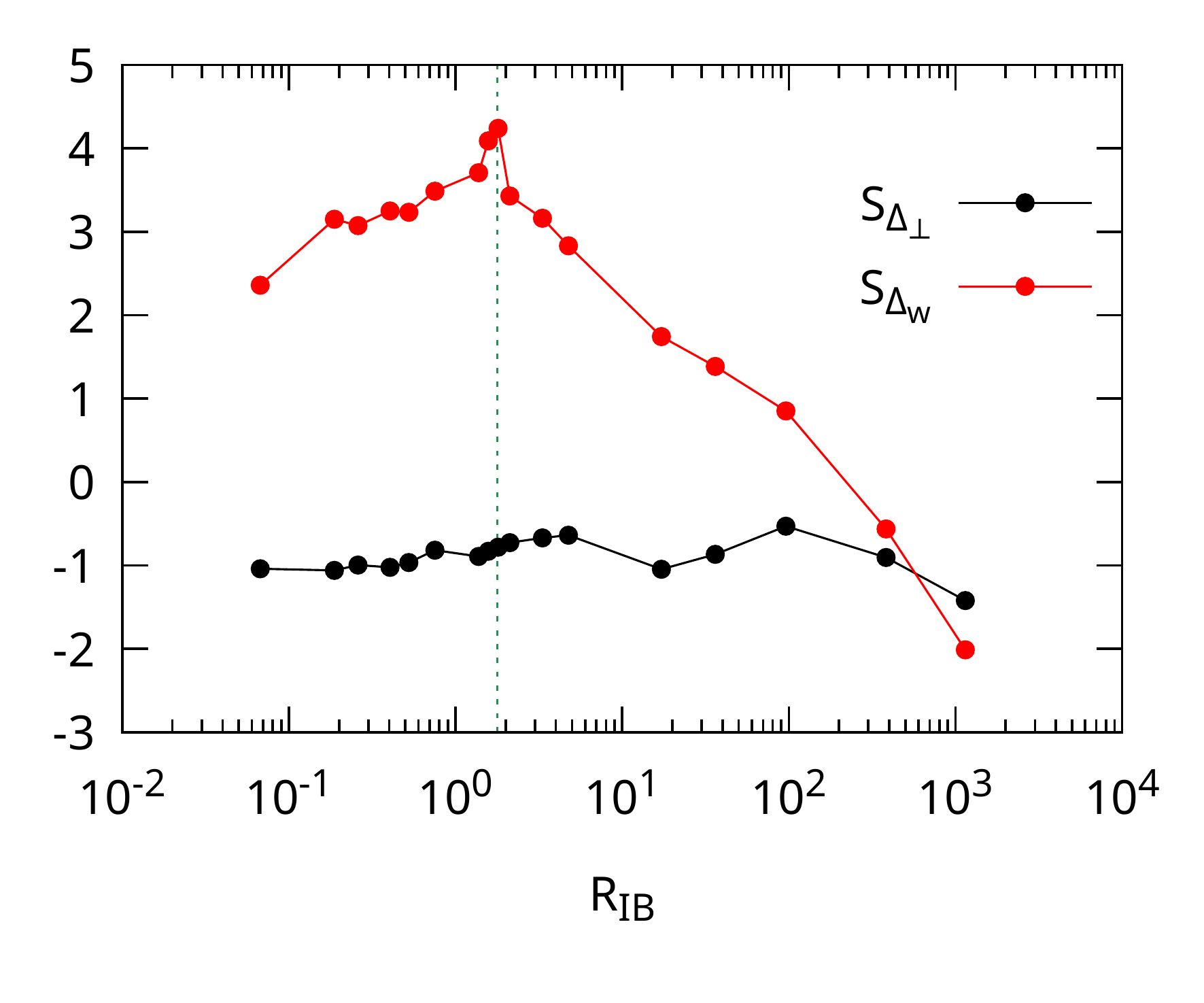}   \includegraphics[width=8.1cm]{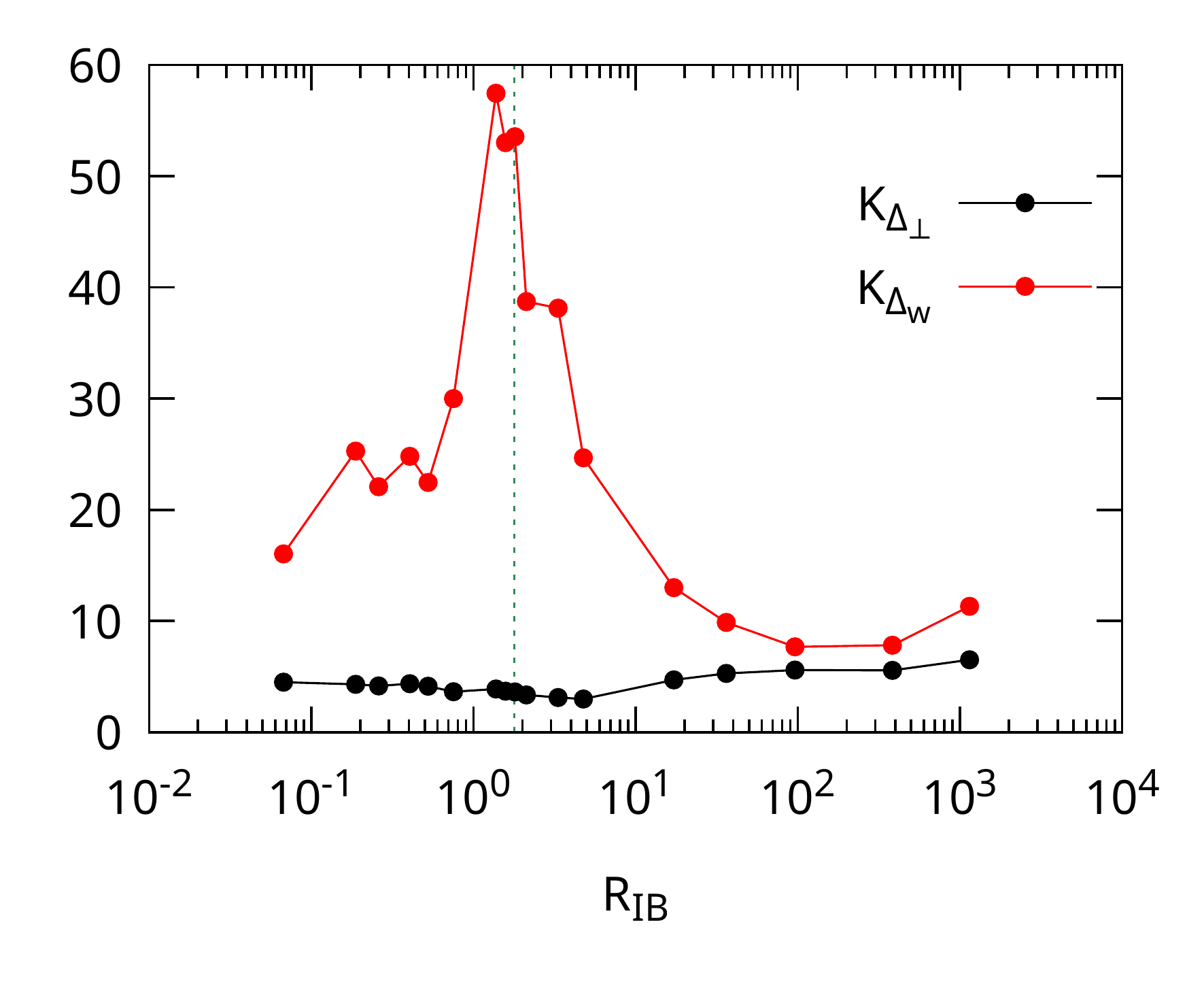} 
\caption{Skewness $S$ (left) and kurtosis $K$ (right) of the normalised perpendicular ($\perp$, black) and vertical ($w$, red) energy defects $\Delta_{\perp,w}$ when contrasted with the potential energy (see eq. \ref{eq:delta}), as a function of buoyancy Reynolds number $R_{IB}$. 
The statistics of the energy difference between the horizontal and potential modes (black curves) do not vary substantially across $R_{IB}$ values.
On the other hand, there is a clear correlation between the lack of equipartition of vertical and potential energy, with strong $[S,K]$ (red plots) and the strong intermittency close to $R_{IB}\approx 1.78$ indicated by the vertical dotted line, particularly visible for the kurtosis given at right.}
 \label{fig:10} \end{figure*} 

Finally, in Fig.~\ref{fig:10} we display the skewness $S$ (left) and kurtosis $K$ (right) of the energy defects, as defined below, for runs of the PP-Pr1 series: 
\be \Delta_{E,\perp}=\frac{E_P-E_{u_\perp}}{E_P + E_V} \ , \ \Delta_{E,w}=\frac{E_P-E_w}{E_P + E_V}
\label{eq:delta} \ee 
where $\Delta_{E,\perp}$ is shown in black and $\Delta_{E,w}$ in red (the two PP-Pr1 runs at very high $R_{IB}$, marked * in Table~\ref{Tab1}, are here omitted for the sake of clarity). 
Both quantities are normalized by the total energy $E_P+E_V$, with $E_P$ and $E_V$ the potential and kinetic energies, and $E_{u_\perp}$ and $E_w$ the energy associated with the horizontal and vertical components of the velocity, all integrated over the domain volume and time. For both $S$ and $K$, there is little variation with $R_{IB}$ for the horizontal energy defect, but a clear change, with a rapid decrease, around $R_{IB}=1.78$, 
for $\Delta_{E,w}$. This does strengthen the argument made before on the basis of the model for intermittency that it is the imbalance between vertical energy and potential energy that feeds the large-scale intermittency.

%

\section{Discussion and conclusion} \label{S:CONCLU}
Strongly stratified flows, by definition, display an imbalance in time scales, between the fast waves and the slow nonlinear eddies. However, when an instability is triggered (convective or Kelvin-Helmoltz for example), small-scales are formed rapidly and dissipate energy efficiently, taking the flow back to a wave-dominated state. In these spatiotemporal bursts \cite{marino_22}, strong intermittency takes place, as diagnosed through the skewness and kurtosis of many variables
($w, \theta, B_f$, see Fig. \ref{fig:1}) \cite{feraco_18,pouquet_19p,feraco_21}; this is a signature of large-scale intermittent events, as observed in stratified geophysical flows. As shown in several studies, the local gradient Richardson number is close to a critical value in large parts of the flow, maintaining the flow to the brink of an instability (see e.g. \cite{salehipour_18, smyth_19, pouquet_19p, smyth_20, sujovolsky_20} and references therein).

Denoting $\Omega$ the rotation frequency, $\approx 10^{-4}s^{-1}$ for Earth at mid latitudes, $N/f$ (where $f=2\Omega$) is of order $10$ or less in the ocean, and $100$ or more in the atmosphere. Strong rotation at small Rossby numbers, with 
$Ro=U_{rms}/[f L_{int}]$, can play a role as well in the development of large-scale and small-scale intermittency. 
It will thus be of interest to study the effect of rotation on large-scale intermittency in the framework of a continuation of the present study, 
 as we already know that, in the decaying case and with quasi-geostrophic initial conditions, a peak in $K_w$ is present when looking at variations of the Eulerian vertical velocity with $R_{IB}$ (and thus for small Froude numbers, see Figure 4c in Pouquet et al.~\cite{pouquet_19p}).
 
However, such computations will prove costly since, in the presence of forcing, an inverse cascade of energy develops at large scale for small Rossby numbers together with a direct energy cascade, both with constant flux \cite{pouquet_13p,marino_15p, Alexakis_2024}.One may therefore need to introduce a large-scale friction term to prevent energy accumulation at the gravest mode when forcing is applied at intermediate scales and simulations are performed for long times. The rate of this large-scale energy buildup, as well as whether it occurs in the parallel and/or perpendicular directions in Fourier space, has been shown to depend on the ratio $N/f$ and on the domain aspect ratio \cite{marino_13i,marino_14,Alexakis_2026}. This is left for future work.

By evaluating both the mixing efficiency ($\langle B_f \rangle /\langle \varepsilon_V \rangle$) and the irreversible mixing efficiency ($\hat \Gamma=\langle \varepsilon_P \rangle / \langle \varepsilon_V \rangle$)  of the flows under study, we note that, in the narrow range of the buoyancy Reynolds number (or Froude number) around the peak of vertical velocity kurtosis, they decrease as the stratification gets stronger until they saturate at the lowest $R_{IB}$, as seen in Figure \ref{fig:7}. From this observation, we see that the extreme events that develop in the flow (both in $w$ and in $\theta$) do not affect globally the mixing efficiency, when this is computed over the flow volume and in time, as is the case as well for other global quantities (see Figs. \ref{fig:5} and \ref{fig:7}).
However, by evaluating the local value of the buoyancy flux $B_f$, we observe that there is indeed a difference in the regions where the extreme events are found. In fact, the contour plots in Figs.~\ref{fig:8} and \ref{fig:9} show that the run with intermediate $R_{IB}$ (middle panel) has stronger fluctuations in the buoyancy flux, which are found for both $w$ and $\theta$. The difference in $B_f$ in this case can be up to one order of magnitude between the regions where both $w$ and $\theta$ develop extreme events and the ones where they do not. 
In the case of stratified turbulence forced by a large-scale shear, it can be shown using a large parametric study that the precise way the turbulence is produced at large scale can influence the outcome in terms of mixing efficiency \cite{yi_23}, a significant point for example in the context of climate parametrizations.

Differences in the local value of the buoyancy flux may be relevant for geophysical applications and for modeling, since we observe this large-scale intermittency in a range of Froude and buoyancy Reynolds numbers that are pertinent for both the atmosphere and the oceans. For example, it is known that the rapid intensification phase of a hurricane is better predicted by following the variation of vertical velocity \cite{mcfarquhar_12, onderlinde_14}. Having identified a new scaling law for the high-order statistics of the buoyancy flux $B_f$ is also very useful for improving the predictability of stratified fluid dynamics in nature. Thus, understanding the conditions that trigger the extreme events characterized in this and previous numerical studies, and observed in the atmosphere and the oceans, is essential for improving the modeling and parametrization of the processes governing the dynamics of geophysical and climatological flows. 

\vskip0.12truein
\noindent {\bf Acknowledgements} \\
All authors contributed equally to this work, and the authors declare no conflict of interest. The computing resources utilized in this work were provided by PMCS2I at the \'Ecole Centrale de Lyon and PSMN at the \'Ecole Normale Superieure de Lyon, in France.
R.M., G.S. acknowledge support from the project ``EVENTFUL'' (no.~ANR-20-CE30-0011) funded by the French Agence Nationale de la Recherche (ANR), and ``STREAM'' (no. APR-2025) funded by the French Centre National d’´Etudes Spatiales (CNES).
F.F. acknowledges the funding from the ASI project “Attività a di Fase A per la missione Plasma Observatory” (2024-15-HH.0). This study was carried out within the Space It Up project funded by the Italian Space Agency, ASI, and the Ministry of University and Research, MUR, under Contract No. 2024-5-E.0-CUP No. I53D24000060005.
A.P. is thankful to Bob Ergun. NCAR is funded in part by NSF.
F.F. and L.P. acknowledge that some of the simulations and part of the analyses have been carried out on the ``Newton" cluster of the High Performance Computing Center of the University of Calabria, supported by EU FP7 2007-13 through the MATERIA Project (PONa3 00370), and EU Horizon 2020 through the STAR 2 Project (PON RI 2014-20, PIR0100008).
P.D.M. acknowledges support from  the project REMATE of Redes Federales de Alto Impacto, Argentina.
\bibliographystyle{apsrev4-1} \bibliography{ap_4Mar26} 

 \end{document}